\documentstyle[12pt]{article}

\def\thefootnote{\fnsymbol{footnote}}
\def\ref#1{$^{#1)}$}
\def\journal{\topmargin .3in    \oddsidemargin .5in
        \headheight 0pt \headsep 0pt
        \textwidth 5.625in 
        \textheight 8.25in 
        \marginparwidth 1.5in
        \parindent 2em
        \parskip .5ex plus .1ex         \jot = 1.5ex}
\journal

\catcode`\@=11
\def\marginnote#1{}
\newcount\hour
\newcount\minute
\newtoks\amorpm
\hour=\time\divide\hour by60
\minute=\time{\multiply\hour by60 \global\advance\minute by-\hour}
\edef\standardtime{{\ifnum\hour<12 \global\amorpm={am}%
        \else\global\amorpm={pm}\advance\hour by-12 \fi
        \ifnum\hour=0 \hour=12 \fi
        \number\hour:\ifnum\minute<10 0\fi\number\minute\the\amorpm}}
\edef\militarytime{\number\hour:\ifnum\minute<10 0\fi\number\minute}
\def\draftlabel#1{{\@bsphack\if@filesw {\let\thepage\relax
   \xdef\@gtempa{\write\@auxout{\string
      \newlabel{#1}{{\@currentlabel}{\thepage}}}}}\@gtempa
   \if@nobreak \ifvmode\nobreak\fi\fi\fi\@esphack}
        \gdef\@eqnlabel{#1}}
\def\@eqnlabel{}
\def\@vacuum{}
\def\draftmarginnote#1{\marginpar{\raggedright\scriptsize\tt#1}}
\def\draft{\oddsidemargin -.5truein
        \def\@oddfoot{\sl preliminary draft \hfil
        \rm\thepage\hfil\sl\today\quad\militarytime}
        \let\@evenfoot\@oddfoot \overfullrule 3pt
        \let\label=\draftlabel
        \let\marginnote=\draftmarginnote
   \def\@eqnnum{(\theequation)\rlap{\kern\marginparsep\tt\@eqnlabel}%
\global\let\@eqnlabel\@vacuum}  }
\def\preprint{\twocolumn\sloppy\flushbottom\parindent 2em
        \leftmargini 2em\leftmarginv .5em\leftmarginvi .5em
        \oddsidemargin -.5in    \evensidemargin -.5in
        \columnsep .4in \footheight 0pt
        \textwidth 10in \topmargin  -.4in
        \headheight 12pt \topskip .4in
        \textheight 7.1in \footskip 0pt
        \def\@oddhead{\thepage\hfil\addtocounter{page}{1}\thepage}
        \let\@evenhead\@oddhead \def\@oddfoot{} \def\@evenfoot{} }
\def\numberbysection{\@addtoreset{equation}{section}
        \def\theequation{\thesection.\arabic{equation}}}
\def\underline#1{\relax\ifmmode\@@underline#1\else
        $\@@underline{\hbox{#1}}$\relax\fi}
\def\titlepage{\@restonecolfalse\if@twocolumn\@restonecoltrue\onecolumn
     \else \newpage \fi \thispagestyle{empty}\c@page\z@
        \def\thefootnote{\fnsymbol{footnote}} }
\def\endtitlepage{\if@restonecol\twocolumn \else \newpage \fi
        \def\thefootnote{\arabic{footnote}}
        \setcounter{footnote}{0}}  
\catcode`@=12
\relax
\def\figcap{\section*{Figure Captions\markboth
        {FIGURECAPTIONS}{FIGURECAPTIONS}}\list
        {Figure \arabic{enumi}:\hfill}{\settowidth\labelwidth{Figure 999:}
        \leftmargin\labelwidth
        \advance\leftmargin\labelsep\usecounter{enumi}}}
 \relax
\def\tablecap{\section*{Table Captions\markboth
        {TABLECAPTIONS}{TABLECAPTIONS}}\list
        {Table \arabic{enumi}:\hfill}{\settowidth\labelwidth{Table 999:}
        \leftmargin\labelwidth
        \advance\leftmargin\labelsep\usecounter{enumi}}}
 \relax
\def\reflist{\section*{References\markboth
        {REFLIST}{REFLIST}}\list
        {[\arabic{enumi}]\hfill}{\settowidth\labelwidth{[999]}
        \leftmargin\labelwidth
        \advance\leftmargin\labelsep\usecounter{enumi}}}
 \relax
\makeatletter
\newcounter{pubctr}
\def\publist{\@ifnextchar[{\@publist}{\@@publist}}
\def\@publist[#1]{\list
        {[\arabic{pubctr}]\hfill}{\settowidth\labelwidth{[999]}
        \leftmargin\labelwidth
        \advance\leftmargin\labelsep
        \@nmbrlisttrue\def\@listctr{pubctr}
        \setcounter{pubctr}{#1}\addtocounter{pubctr}{-1}}}
\def\@@publist{\list
        {[\arabic{pubctr}]\hfill}{\settowidth\labelwidth{[999]}
        \leftmargin\labelwidth
        \advance\leftmargin\labelsep
        \@nmbrlisttrue\def\@listctr{pubctr}}}
 \relax
\makeatother
\catcode`\@=11
\def\section{\@startsection {section}{1}{0pt}{-3.5ex plus -1ex minus
 -.2ex}{2.3ex plus .2ex}{\raggedright\large\bf}}
\catcode`\@=12
\newskip\humongous \humongous=0pt plus 1000pt minus 1000pt

\newif\ifdtup

\def\oldreffmt#1{\rlap{[#1]} \hbox to 2\parindent{}}

\def\figfmt#1{\rlap{Figure {#1}} \hbox to 1in{}}

\def\beq{\begin{equation}}
\def\eeq{\end{equation}}

\def\bea{\begin{eqnarray}}

\def\eea{\end{eqnarray}}
\catcode`\@=11
\def\eqnarray{\stepcounter{equation}\let\@currentlabel=\theequation
\global\@eqnswtrue
\global\@eqcnt\z@\tabskip\@centering\let\\=\@eqncr
\gdef\@@fix{}\def\eqno##1{\gdef\@@fix{##1}}%
$$\halign to \displaywidth\bgroup\@eqnsel\hskip\@centering
  $\displaystyle\tabskip\z@{##}$&\global\@eqcnt\@ne
  \hskip 2\arraycolsep \hfil${##}$\hfil
  &\global\@eqcnt\tw@ \hskip 2\arraycolsep $\displaystyle\tabskip\z@{##}$\hfil
   \tabskip\@centering&\llap{##}\tabskip\z@\cr}

\def\@@eqncr{\let\@tempa\relax
    \ifcase\@eqcnt \def\@tempa{& & &}\or \def\@tempa{& &}
      \else \def\@tempa{&}\fi
     \@tempa \if@eqnsw\@eqnnum\stepcounter{equation}\else\@@fix\gdef\@@fix{}\fi
     \global\@eqnswtrue\global\@eqcnt\z@\cr}

\catcode`\@=12
\font\tenbifull=cmmib10 
\font\tenbimed=cmmib10 scaled 800
\font\tenbismall=cmmib10 scaled 666
\relax

\begin{document}

\begin{titlepage}
\begin{center}
\today\     \hfill    LBL-36516 \\
            \hfill    UCB-PTH-94/34 \\
\vskip .5in

{\large \bf The Minimal Supersymmetric Model\\ of \\
Higgs-Higgs Condensation}
\footnote{This work was supported in part by the Director, Office of
Energy Research, Office of High Energy and Nuclear Physics, Division of
High Energy Physics of the U.S. Department of Energy under Contract
DE-AC03-76SF00098 and in part by the National Science Foundation under
grant PHY90-21139.}

\vskip .5in
Yi-Yen Wu\\[.5in]

{\em  Theoretical Physics Group\\
      Lawrence Berkeley Laboratory\\
      University of California\\
      Berkeley, CA 94720}
\end{center}

\vskip .5in

\newpage
\begin{abstract}
    This paper is motivated by three issues associated with
the supersymmetric extension of the standard model:
the \mbox{$\mu$ problem}, the possibility of raising the upper
bound on the lightest-Higgs-boson mass, and the triviality problem
associated with the Higgs sector. A new scheme based on the
Higgs-Higgs condensation at low energy is proposed,
and it is shown that these three issues are well solved by
this scheme. As the first realization of this new scheme,
the Minimal Supersymmetric Model of Higgs-Higgs Condensation
(MSMHHC) is constructed and studied in detail. The MSMHHC is
identical with the MSSM (Minimal Supersymmetric Standard Model) in
the fundamental particle content, and their lagrangians differ
only in the Higgs sector. The Higgs sector of the MSMHHC
is based on the softly-broken supersymmetric Nambu$-$Jona-Lasinio
model with the four-field interaction of the Higgs doublets.
At low energy, these two Higgs doublets condense into
two neutral Higgs gauge singlets, and the low-energy effective
lagrangian of the MSMHHC has the form of the non-minimal
supersymmetric standard model which contains two more
Higgs singlets than the MSSM. Another unique feature of the MSMHHC is that
heavy top quark always implies strongly-interacting
low-energy Higgs sector, i.e., a large mass for the lightest
Higgs boson. A systematic study of the parameter space is also made in
order to reveal the qualitative features of the MSMHHC.
Finally, we comment on the question
{\it "How large can the lightest-Higgs mass be?"} with the conclusion
that the supersymmetric model of Higgs-Higgs condensation will be the most
promising candidate if the lightest-Higgs mass of the MSSM is excluded by
the future experiments.
\end{abstract}
\end{titlepage}
\renewcommand{\thepage}{\roman{page}}
\setcounter{page}{2}
\mbox{ }

\vskip 1in

\begin{center}
{\bf Disclaimer}
\end{center}

\vskip .2in

\begin{scriptsize}
\begin{quotation}
This document was prepared as an account of work sponsored by the United
States Government.  Neither the United States Government nor any agency
thereof, nor The Regents of the University of California, nor any of their
employees, makes any warranty, express or implied, or assumes any legal
liability or responsibility for the accuracy, completeness, or usefulness
of any information, apparatus, product, or process disclosed, or represents
that its use would not infringe privately owned rights.  Reference herein
to any specific commercial products process, or service by its trade name,
trademark, manufacturer, or otherwise, does not necessarily constitute or
imply its endorsement, recommendation, or favoring by the United States
Government or any agency thereof, or The Regents of the University of
California.  The views and opinions of authors expressed herein do not
necessarily state or reflect those of the United States Government or any
agency thereof of The Regents of the University of California and shall
not be used for advertising or product endorsement purposes.
\end{quotation}
\end{scriptsize}

\vskip 2in

\begin{center}
\begin{small}
{\it Lawrence Berkeley Laboratory is an equal opportunity employer.}
\end{small}
\end{center}

\newpage
\renewcommand{\thepage}{\arabic{page}}
\setcounter{page}{1}
\section{Introduction}
\hspace{0.8cm}The Minimal Supersymmetric Standard Model (MSSM)
is by far the most studied supersymmetric extension of the
standard model. However, there are still unsolved issues
about the MSSM, and it may be necessary to go beyond the MSSM for a
better solution. In this paper, we will confine ourselves to the issues
of the Higgs sector. The first issue is the $\mu$-term
problem, $\mu (H_{1}\epsilon H_{2})$, of the MSSM \cite{mu-termN,mu-termH},
where $H_{1}$ and $H_{2}$ are the two Higgs-doublet superfields.
In the MSSM, we need $\mu$ to be non-zero
and of order the weak scale. However, since $\mu (H_{1}\epsilon H_{2})$
is supersymmetric and gauge invariant, there is no reason for
$\mu$ not to be of order the Planck scale or the GUT scale. Thus,
there is really a naturalness problem of $\mu$ in the MSSM and ideas
beyond the MSSM are needed to solve this problem in a natural
way. The second issue is about the upper bound on the
lightest-Higgs-boson mass. It is well-known that, in the MSSM, the
lightest-Higgs mass at tree level cannot get larger than the mass of
gauge boson $Z$, and this upper bound is raised as much as
\mbox{20 (50) GeV} for a top-quark mass of \mbox{150 (200) GeV}
\cite{Haber91,Haber92}. However, there is no guarantee of
finding a light Higgs below the upper bound of the MSSM in
future experiments. It is then of essential importance to
go beyond the MSSM and consider those models which can
raise the upper bound of the lightest-Higgs mass above
the prediction of the MSSM naturally. One of those existing models is the
extension of the MSSM by including more Higgs gauge singlets in the Higgs
sector and more non-trivial Higgs self-interactions in the
superpotential (i.e., the Non-Minimal Supersymmetric Standard Model,
\mbox{Non-Minimal SSM} \cite{HiggsHunter}). In the Non-Minimal SSM,
a large mass for the lightest Higgs boson always implies a
strongly-interacting Higgs sector. And, a strongly-interacting
Higgs sector suffers from the triviality problem associated with
the Higgs self-couplings \cite{Callaway,Binetruy92,yywu}.
Therefore, the third issue is how this triviality problem
can be resolved at high energy.
The above three issues actually constitute the main motivations of
this paper, and it turns out that the Minimal Supersymmetric Model
of Higgs-Higgs Condensation (MSMHHC) proposed here
does provide the natural answers to all these issues.
Next, we will briefly describe the existing approaches
and our approach to these issues.

    Let's begin with the first issue, the \mbox{$\mu$ problem}. There
have been two attitudes towards the natural $\mu$-term: either derive
the $\mu$-term directly from the high-energy fundamental theory
(e.g., the superstring or supergravity) or generate the $\mu$-term
within the low-energy (i.e., lower than the Planck scale) effective
models. The latter attitude is adopted in this paper and, since those
works which adopt the former attitude are irrelevant to our
discussions of the second and the third issues, the attention will
be focused only on those works which adopt the latter attitude. A
direct step beyond the MSSM is the Next to Minimal Supersymmetric Standard
Model (NMSSM or (M+1)SSM) \cite{NMSSM}, which has one Higgs gauge
singlet $N$ and two Higgs doublets $H_{1}$, $H_{2}$.
The relevant piece in the superpotential $W_{NMSSM}$ is
$\,\,hN(H_{1}\epsilon H_{2})$, where $h$ is the Higgs coupling. In the NMSSM,
the $\mu$-term is effectively generated through
$\,\mu=h<\phi_{N}>$, where $\phi_{N}$ is the scalar
component of $N$. However, the naturalness problem of $\mu$
remains unsolved in the NMSSM because there is again the naturalness
problem of $<\phi_{N}>$. Another effective approach is the
Minimal Top-Condensate Supersymmetric Standard Model (MTCSSM)
\cite{Clark90,Carena92,Binetruy94,Buchmuller84}, where there is
no fundamental Higgs field and the Higgs at low
energy is actually the top condensate. Starting with the
Supersymmetric Nambu$-$Jona-Lasinio (SUSY NJL) model with the
four-field interaction of the top defined at the cut-off, the low-energy
effective lagrangian of the MTCSSM always contains the $\mu$-term.
It was pointed out in \cite{Binetruy94} that $\mu$ is naturally
of order the soft SUSY-breaking scalar mass. Hence, $\mu$ is
naturally small and the MTCSSM does solve the \mbox{$\mu$ problem}.

    Our approach, the Minimal Supersymmetric Model of Higgs-Higgs
Condensation (MSMHHC), which has exactly the
same fundamental particle content as the MSSM does, is based
on the softly-broken SUSY NJL model with the four-field interaction
of two Higgs doublets, $H_{1}$ and $H_{2}$, defined
at the cut-off $\Lambda$.
At low energy, $H_{1}$ and $H_{2}$ condense into two composite
neutral Higgs-singlet superfields, $M$ and $N$ \cite{Binetruy94},
and the superpotential $W_{MSMHHC}$ of the low-energy effective
lagrangian has the form of the Non-Minimal SSM:
\mbox{$W_{MSMHHC}\,=\,mMN\,+\,hN(H_{1}\epsilon H_{2})$} (the soft
SUSY-breaking terms are not written down here), where $m$
and $h$ are coupling constants. So, the MSMHHC looks like the MSSM at high
energy, but behaves like the Non-Minimal SSM with two Higgs singlets
(which we call the (M+2)SSM) at low energy.
Obviously, the MSMHHC generates the $\mu$-term in
the same way as the NMSSM does, where $\,\mu=h<\phi_{N}>$. Nevertheless,
unlike the NMSSM, $<\phi_{N}>$ is generated dynamically in the MSMHHC
due to supersymmetry breaking and naturally of order the soft
SUSY-breaking scalar mass. That is, $<\phi_{N}>$ vanishes in the
supersymmetric limit. Therefore, $\mu$ is naturally small
because $\,h\,\approx\,$O(1) naturally in the MSMHHC. The MSMHHC does provide
a natural solution to the \mbox{$\mu$ problem}. This is indeed
an explicit realization of the suggestion in \cite{mu-termH}:
$\mu$ is zero initially, and a non-zero value is induced only
through supersymmetry breaking effects.

    In fact, there will be another potential disadvantage of the NMSSM
(or, any Non-Minimal SSM containing fundamental
neutral Higgs singlets) if the fundamental
Higgs singlet is introduced to generate the
$\mu$-term. Assuming certain SUSY grand unification schemes,
the Higgs singlet coupled to the Higgs doublets at
low energy may destroy the hierarchy between the light fields
and the superheavy fields if the Higgs singlet is also
coupled to the superheavy fields \cite{Polchinski82}.
However, in the MSMHHC, there is no fundamental Higgs singlet
and the Higgs singlets at low energy are composite.
Therefore, in the MSMHHC, there is no danger of destroying the
hierarchy between the light and the superheavy fields. From this
viewpoint, the MSMHHC shares the advantages of both the MSSM
at high energy and the Non-Minimal SSM at
low energy, but avoids the disadvantages of both the MSSM at
low energy and the Non-Minimal SSM at high energy. Even as
a purely theoretical construction, the MSMHHC is worth studying
because it provides us with a interesting dynamic
connection between the MSSM at high energy and the
\mbox{Non-Minimal SSM} at low energy.

    For the sake of completeness, we also mention those
works on the \mbox{$\mu$ problem} in the context of supergravity
or superstring theories briefly, although they are not really
relevant to our approach. Based on Peccei-Quinn symmetry
\cite{PQ}, the \mbox{$\mu$ problem} and the
strong-CP problem can be combined,
and the $\mu$-term can be generated by a
composite axion in the hidden sector \cite{Kim91,Chun92} or by
a generalized Higgs mass term \cite{Chun94} naturally. There are other
approaches without Peccei-Quinn symmetry \cite{Guidice88}.
Notice that  \cite{Kim91,Chun92} is the
first work that proposed the idea of compositeness in solving
the $\mu$ problem. However, in  \cite{Kim91,Chun92}, the
$\mu$-term is generated by coupling the hidden-sector superfields
$S_{1}$, $S_{2}$ to the Higgs superfields as
\mbox{$\frac{1}{M} S_{1} S_{2} (H_{1}\epsilon H_{2})$}
\mbox{($M\,\simeq\, 2.43\times 10^{18}$ GeV)},
and the scalar components of $S_{1}$,
$S_{2}$ condense into the composite axion in the hidden
sector. Hence, the scheme and the main results of our model MSMHHC,
which is based on the Higgs-Higgs condensation, are completely
different from those of \cite{Kim91,Chun92}.

    Now, turn to the second and the third issues. In general,
in the \mbox{Non-Minimal SSM}, the upper bound on the
lightest-Higgs mass is expected to be relaxed because the tree-level
constraint on the lightest-Higgs mass of the MSSM is no longer valid.
For example, in the NMSSM,
$\,m_{LH} \leq \frac{1}{\sqrt{2}}hv\,$
(when $\tan\beta$=1) has been derived in \cite{Binetruy92}, where
$v\approx$250 GeV. $m_{LH}$ denotes the physical mass of the lightest Higgs
boson throughout this paper.
Increasing the Higgs coupling $h$ therefore
increases the upper bound. However, as
mentioned before, one then has to face the triviality problem.
That is, when scaled toward high energy, $h$ will blow up
faster if a larger value of $h$ is assumed at low energy.
Based on the observation of triviality, several estimates
of the upper bound of $m_{LH}$ in the NMSSM have been made
\cite{Binetruy92,yywu,Moroi92}, and they are indeed larger
than that of MSSM. To go beyond the triviality
problem, it has been argued in \cite{Lindner93} that the problem
of triviality is possibly an indication of compositeness. From
this viewpoint, the MSMHHC is precisely the realization of the argument
of \cite{Lindner93} applied to the triviality problem in the
\mbox{Non-Minimal SSM}. In the MSMHHC, the composite field $N$ is
static at tree level, but develops a kinetic term at
low energy, $Z_{N} \int d^{4}\theta \,N^{\dagger}N$, from
quantum corrections \cite{Carena92}, where
\mbox{$Z_{N} \sim \frac{1}{8\pi^{2}}\ln (\frac{\Lambda^{2}}{\mu_{E}^{2}})$}
is the wave-function renormalization and $\mu_{E}$ is the
renormalization scale. The superpotential of the low-energy
effective lagrangian is $W_{MSMHHC}=\,mMN\,+\,hN(H_{1}\epsilon H_{2})$,
and $\,h \sim \frac{1}{\sqrt{Z_{N}}}$ because we have rescaled $N$
\mbox{($N\,\rightarrow\,\frac{1}{\sqrt{Z_{N}}}N$)} in order to have the
correct normalization of the kinetic term in the low-energy
effective lagrangian. In the limit
\mbox{$\mu_{E}\,\rightarrow\,\Lambda$}, the kinetic term of $N$
vanishes (i.e., $N$ is then static) and $h$ blows up. Therefore,
the triviality problem is resolved within the MSMHHC in the sense that
the triviality problem seems to arise only because $N$ is mistaken
for a dynamical field near the cut-off $\Lambda$. The MSMHHC, which
looks like the \mbox{Non-Minimal SSM} at low energy
but without the plague of triviality, is a promising answer
to the last two issues. In \mbox{Sections 5$-$6}, it will
further be illustrated that the large value of $m_{LH}$ predicted
by the MSMHHC is closely related to the strong
\mbox{top Yukawa interaction}.

    The MTCSSM is irrelevant to the above discussion because the
low-energy form of the MTCSSM is simply the MSSM and there is no Higgs
self-interaction. Although the MTCSSM can also solve the
\mbox{$\mu$ problem}, the MSMHHC solves it with a scheme
different from that of the MTCSSM, and therefore is worth studying.
Besides, the MSMHHC (whose low-energy theory is the (M+2)SSM)
is more interesting than the MTCSSM (whose low-energy theory
is the MSSM) from the viewpoint of raising the upper
bound on the lightest Higgs mass. The present version
of the MSMHHC contains not only the Higgs sector
but also the \mbox{top Yukawa term}. It is reasonable to neglect
the other quark and lepton Yukawa couplings for a qualitative
description of the phenomena under study. The bottom Yukawa coupling
is also neglected because the possibility of \mbox{$\,\tan\beta\gg 1\,$}
is not considered in this paper. In the following, we shall briefly
describe the organization of this paper.

    In Section 2, the MSMHHC is defined and its low-energy
effective lagrangian is derived using the SUSY NJL technique.
The dynamics of the composite Higgs singlets is discussed,
and the triviality problem is solved naturally. In
\mbox{Section 3}, in order to study the MSMHHC vacuum and related
physical quantities, the effective potential of the MSMHHC is computed.
In \mbox{Section 4}, from the minimization of the effective potential,
the vacuum constraint equations are derived and the MSMHHC vacuum is
examined. In the MSMHHC, the spontaneous breaking of
electroweak symmetry accompanies the condensation of Higgs
doublets, and therefore, compared to the conventional NJL model, the MSMHHC
has two more vacuum constraints. In \mbox{Section 5}, the full
mass spectrum of the MSMHHC is analyzed. A unique feature of the MSMHHC is
that the lightest Higgs boson will be massless if the \mbox{top Yukawa
interaction} is turned off, and the lightest Higgs boson
becomes massive only through the effects of the
\mbox{top Yukawa interaction}. It is then shown
that, in the MSMHHC, strong \mbox{top Yukawa interaction}
(i.e., heavy top quark) usually implies a large mass for
the lightest Higgs boson. The same conclusion will also be
reached elsewhere in this paper.

    In \mbox{Section 6}, to study the phenomenological aspects of the MSMHHC
in general, its full parameter space is defined, and its behavior
over the parameter space is examined in Sections 6$-$9. All the
physical quantities are first computed from the effective
potential at the cut-off scale, and
then renormalized down to low energy properly. Only two
phenomenologically reasonable constraints are assumed in the analysis of
the parameter space. The first constraint requires that all the soft
SUSY-breaking scalar squared masses in the Higgs and the top sectors be
non-negative at low energy, and this leads to two non-trivial
results. First, this constraint implies a simple relationship among
the soft SUSY-breaking scalar masses. Second, under this constraint,
strong \mbox{top Yukawa interaction} always implies
strongly-interacting low-energy Higgs sector
(i.e., large $m_{LH}$). The conclusion of
\mbox{Section 5} is re-established in a different way, and
therefore it is indeed a consistent feature of MSMHHC. The second
constraint requires the physical top-quark mass to be
\mbox{164$\,\sim\,$180 GeV}, and, together with the first constraint,
it implies that the MSMHHC should be an effective intermediate-scale
model, i.e., \mbox{weak scale $\ll\,\Lambda\,\ll$ Planck scale}.
However, there is a fine-tuning problem associated with the soft
SUSY-breaking parameters (although it's not severe).
It is argued that this fine-tuning problem
is actually a guide to the future model-building, not a
real obstacle. In \mbox{Sections 7$-$9}, the low-energy physical
quantities, such as the mass of the lightest Higgs $m_{LH}$,
$\tan\beta_{r}$, $\mu_{r}$, etc., are
computed and their dependence on the parameter space is examined
systematically. In \mbox{Section 7}, the dependence on the cut-off
$\Lambda$ is studied. In one example, the MSMHHC
predicts \mbox{150 GeV $<\,m_{LH}\,<$ 400 GeV}. The fact
that $m_{LH}$ can be as large as \mbox{400 GeV} indeed
solves the second issue. In \mbox{Section 8}, the dependence on
$F_{SUSY}$ (the strength of soft SUSY breaking) is
studied. As an explicit answer to the \mbox{$\mu$ problem}, the
dependence of the effective $\mu_{r}$ on $F_{SUSY}$ is
emphasized. Another important phenomenon is the saturation of
$m_{LH}$ in the limit of large $F_{SUSY}$. In
\mbox{Section 9}, the dependence on the soft SUSY-breaking pattern
is studied, and  we are especially interested in how $\tan\beta_{r}$ and
$m_{LH}$ depend on the soft SUSY-breaking pattern.
Finally, it is pointed out in \mbox{Section 10} that, in order to
make the phenomenological study of the MSMHHC more complete, direct
extensions of the present model are necessary and guidelines
for the future model-building are needed. We also comment on the interesting
question: {\it "How large can the lightest-Higgs mass be?"}
\section{The Minimal Supersymmetric Model of
         \mbox{Higgs-Higgs Condensation}}
\hspace{0.8cm} The Minimal Supersymmetric Model of Higgs-Higgs
Condensation (MSMHHC) is minimal in the sense that the MSMHHC
and the MSSM are identical in the fundamental
particle content, and their lagrangians differ only in the Higgs
sector. Viewing this work as a first attempt at the idea of Higgs-Higgs
condensation, we choose to study the main physical features of
the MSMHHC rather than make it a phenomenologically complete model.
Therefore, reasonable simplifications will be made whenever it is
necessary. We shall ignore all quark and lepton Yukawa couplings
except the one associated with the top quark, since the others are
inessential to the qualitative description of the phenomena under
study. Notice that the bottom Yukawa coupling is also neglected because
the possibility of \mbox{$\tan\beta\gg 1$} is not considered in
this paper. The SU(3)$_{c}$ color symmetry is ignored because
we are mainly concerned with the scalar Higgs sector. We then start
with the following lagrangian of the MSMHHC, $\cal L$$_{\Lambda}$, which is a
non-renormalizable model with momentum cut-off $\Lambda$:
\begin{eqnarray}
{\cal L}_{\Lambda}&=&\int d^{4}\theta\,
           \{\,H_{1}^{\dagger}H_{1}(1-m_{1}^{2}\theta^{2}\bar{\theta}^{2})+
             H_{2}^{\dagger}H_{2}(1-m_{2}^{2}\theta^{2}\bar{\theta}^{2})
                                                          \nonumber \\
& &\hspace{1.5in}+(Q^{\dagger}Q+T_{C}^{\dagger}T_{C})
              (1-m_{T}^{2}\theta^{2}\bar{\theta}^{2}) \,\}\; \nonumber \\
           & &+\int d^{4}\theta\,G(H_{1}\epsilon H_{2})^{\dagger}
                                 (H_{1}\epsilon H_{2})
              [1+B\theta^{2}+B\bar{\theta}^{2}+
               (B^{2}-m_{M}^{2})\theta^{2}\bar{\theta}^{2}]\; \nonumber \\
           & &+\int d^{2}\theta\,
              \{m_{12}^{2}(H_{1}\epsilon H_{2})\theta^{2}+
               f_{T}(H_{2}\epsilon Q)T_{C}\}\; \nonumber \\
           & &+\int d^{2}\bar{\theta}\,
              \{m_{12}^{2}(H_{1}\epsilon H_{2})^{\dagger}\bar{\theta}^{2}+
               f_{T}(H_{2}\epsilon Q)^{\dagger}T_{C}^{\dagger}\}
\end{eqnarray}
The MSMHHC has the global symmetry SU(2)$\times$U(1), and its Higgs sector
is based on the softly-broken SUSY NJL model. $H_{1}$ and $H_{2}$ are
the two SU(2) Higgs doublets. $Q$ is the SU(2) doublet of the top and
the bottom chiral superfields. $T_{C}$ is the SU(2) singlet
of the top (the SU(2) bottom singlet is omitted since the bottom Yukawa
coupling is neglected). $\epsilon$ is the usual 2$\times$2
antisymmetric $\epsilon$-tensor, and
$(H_{1}\epsilon H_{2})=\epsilon_{ij}H_{1i}H_{2j}$
is implied. The convention for superspace notations of \cite{Wess} is
adopted. $f_{T}$ is the top Yukawa coupling. $G$ is the four-Fermi
coupling constant of dimension mass$^{-2}$.
\mbox{$m_{1}$, $m_{2}$, $m_{T}$, $m_{M}$, $B$ and $m_{12}$}
are the six soft SUSY-breaking parameters of dimension mass.

    To generalize the usual NJL technique to the supersymmetric
case, it was first pointed out in \cite{Buchmuller84,Buchmuller82}
that one has to introduce two chiral superfields (denoted as $M$
and $N$ here) in order to write a linearized version of the
SUSY NJL model. Therefore, with the introduction of $M$ and $N$,
$\cal L$$_{\Lambda}$ can be written in a more instructive form:
\begin{eqnarray}
{\cal L}_{\Lambda}&=&\int d^{4}\theta\,
           \{\,H_{1}^{\dagger}H_{1}(1-m_{1}^{2}\theta^{2}\bar{\theta}^{2})+
             H_{2}^{\dagger}H_{2}(1-m_{2}^{2}\theta^{2}\bar{\theta}^{2})
                                                          \nonumber \\
& &\hspace{.5in}
             +M^{\dagger}M(1-m_{M}^{2}\theta^{2}\bar{\theta}^{2})+
              (Q^{\dagger}Q+T_{C}^{\dagger}T_{C})
              (1-m_{T}^{2}\theta^{2}\bar{\theta}^{2}) \,\}\;
                                                          \nonumber \\
           & &+\int d^{2}\theta\,
              \{m_{12}^{2}(H_{1}\epsilon H_{2})\theta^{2}+
               mMN(1-B\theta^{2})+hN(H_{1}\epsilon H_{2})+
               f_{T}(H_{2}\epsilon Q)T_{C}\}\;
                                                          \nonumber \\
           & &+\int d^{2}\bar{\theta}\,
              \{m_{12}^{2}(H_{1}\epsilon H_{2})^{\dagger}\bar{\theta}^{2}+
               mM^{\dagger}N^{\dagger}(1-B\bar{\theta}^{2})+
               hN^{\dagger}(H_{1}\epsilon H_{2})^{\dagger}+
               f_{T}(H_{2}\epsilon Q)^{\dagger}T_{C}^{\dagger}\}
                                                          \nonumber \\
           & &
\end{eqnarray}
where
\begin{equation}
         G=\frac{h^{2}}{m^{2}}
\end{equation}
and, from (2), the Euler-Lagrange equations of $M$ and $N$ are:
\begin{eqnarray}
M&=&-\frac{h}{m}(H_{1}\epsilon H_{2})(1+B\theta^{2}) \nonumber \\
N&=&\frac{h}{4m^{2}}(1+B\theta^{2})\{ \bar{D}^{2}
     [(H_{1}\epsilon H_{2})^{\dagger}
      (1+B\bar{\theta}^{2}-m_{M}^{2}\theta^{2}\bar{\theta}^{2})] \}
\end{eqnarray}
It is clear from (4) that $M$ and $N$ are indeed
neutral SU(2) singlet composite chiral superfields.
By substituting (4) into (2), (2) is equivalent to the original
lagrangian (1). Before discussing the low-energy effective
lagrangian for (2), let's explain the choice of soft SUSY-breaking
parameters. The set of these six soft SUSY-breaking parameters,
($m_{1}$, $m_{2}$, $m_{T}$, $m_{M}$, $B$, $m_{12}$), is actually
the minimal choice from the viewpoint of dynamical
chiral symmetry breaking and the consideration of the global
SU(2)$\times$U(1) symmetry. It was pointed out in
\cite{Clark90,Carena92,Buchmuller84,Buchmuller82}
that soft SUSY-breaking terms
are necessary in order to get a chiral symmetry breaking vacuum
and to induce condensation. ($m_{1}$, $m_{2}$, $m_{T}$, $m_{M}$)
turns out to be the minimal choice required by the MSMHHC from this
viewpoint. However, without ($B$, $m_{12}$), the Higgs sector of
$\cal L$$_{\Lambda}$ has an SU(2)$\times$U(1)$\times$U(1) symmetry.
Therefore, ($B$, $m_{12}$) is indeed the minimal choice that can explicitly
break this symmetry to the correct electroweak SU(2)$\times$U(1) symmetry.
In general, we could have chosen non-universal soft SUSY-breaking
scalar squared masses at the cut-off for $Q$ and $T_{C}$ in (1). However,
this issue of non-universality is not essential to our main
concerns here.

    In (2), $\cal L$$_{\Lambda}$ looks like the
\mbox{Non-Minimal SSM} which has two more Higgs gauge singlets, $M$
and $N$, than the MSSM does. However, at tree level, $N$ remains static.
When quantum corrections are included, $N$ does develop
a kinetic term \cite{Clark90,Carena92}. According to (2),
there are four divergent (when $\Lambda\rightarrow\infty$)
supergraphs at one loop, and their contributions
can be computed easily:
\begin{eqnarray}
   & &\Sigma_{N}\int d^{4}\theta\,N^{\dagger}N
      [1+(m_{1}^{2}+m_{2}^{2})\theta^{2}\bar{\theta}^{2}],
                                               \nonumber \\
   & &\Sigma_{H_{2}}\int d^{4}\theta\,H_{2}^{\dagger}H_{2}
      [1+2m_{T}^{2}\theta^{2}\bar{\theta}^{2}],
                                               \nonumber \\
   & &\Sigma_{Q}\int d^{4}\theta\,Q^{\dagger}Q
      [1+(m_{2}^{2}+m_{T}^{2})\theta^{2}\bar{\theta}^{2}],
                                               \nonumber \\
   & &\Sigma_{T_{C}}\int d^{4}\theta\,T_{C}^{\dagger}T_{C}
      [1+(m_{2}^{2}+m_{T}^{2})\theta^{2}\bar{\theta}^{2}]
\end{eqnarray}
\begin{eqnarray}
  \Sigma_{N}&=&N_{W}\frac{h^{2}}{16\pi^{2}}\ln
               (\frac{\Lambda^{2}}{\mu_{E}^{2}}), \nonumber \\
  \Sigma_{H_{2}}&=&\Sigma_{Q}=\frac{f_{T}^{2}}{16\pi^{2}}\ln
               (\frac{\Lambda^{2}}{\mu_{E}^{2}}), \nonumber \\
  \Sigma_{T_{C}}&=&N_{W}\frac{f_{T}^{2}}{16\pi^{2}}\ln
               (\frac{\Lambda^{2}}{\mu_{E}^{2}})
\end{eqnarray}
where $\mu_{E}$ is the renormalization scale, and $N_{W}$ is the
dimension of the SU(2) representation. In the present case, $N_{W}$=2.
The first term in (5), which corresponds to the supergraph of Fig.1,
is indeed the kinetic term of $N$. Similar to Fig.1, the other terms
in (5) are generated through the top Yukawa interaction.
The wave-function renormalization constants for these superfields,
in the one-loop approximation, are then defined as follows:
\begin{eqnarray}
  Z_{N}&=&\Sigma_{N},\hspace{.5in}Z_{H_{2}}=1+\Sigma_{H_{2}}, \nonumber \\
  Z_{Q}&=&1+\Sigma_{Q},\hspace{.5in}Z_{T_{C}}=1+\Sigma_{T_{C}}
\end{eqnarray}
$H_{1}$ and $M$ do not get renormalized. The above results show that
these two composite Higgs singlets, $M$ and $N$, are true
dynamical degrees of freedom at low energy ($\mu_{E}\ll\Lambda$),
that is, we should see two Higgs doublets and
two Higgs singlets at low energy. However, as
\mbox{$\mu_{E}\rightarrow\Lambda$}, \mbox{$Z_{N}\rightarrow 0$} and
therefore it no longer makes sense to treat $N$ as
a true dynamic superfield near the cut-off.

    Up to some finite contributions, the low-energy effective
lagrangian $\cal L$$_{eff}$ can be obtained by absorbing
the wave-function renormalization constants and re-defining the
coupling constants according to the results of (5) and (6):
\begin{eqnarray}
\lefteqn{{\cal L}_{eff}=} \nonumber \\
& &      \int d^{4}\theta\,
         \{\,H_{1r}^{\dagger}H_{1r}(1-m_{1r}^{2}\theta^{2}\bar{\theta}^{2})+
             H_{2r}^{\dagger}H_{2r}(1-m_{2r}^{2}\theta^{2}\bar{\theta}^{2})
                                                          \nonumber \\
& &\hspace{.5in}
         +M_{r}^{\dagger}M_{r}(1-m_{Mr}^{2}\theta^{2}\bar{\theta}^{2})+
          N_{r}^{\dagger}N_{r}(1-m_{Nr}^{2}\theta^{2}\bar{\theta}^{2})
                                                         \nonumber \\
& &\hspace{.5in}
         +Q_{r}^{\dagger}Q_{r}(1-m_{Qr}^{2}\theta^{2}\bar{\theta}^{2})+
          T_{Cr}^{\dagger}T_{Cr}(1-m_{T_{Cr}}^{2}\theta^{2}\bar{\theta}^{2})
\,\}\;
                                                          \nonumber \\
           & &+\int d^{2}\theta\,
              \{m_{12r}^{2}(H_{1r}\epsilon H_{2r})\theta^{2}+
    m_{r}M_{r}N_{r}(1-B_{r}\theta^{2})+h_{r}N_{r}(H_{1r}\epsilon H_{2r})+
               f_{Tr}(H_{2r}\epsilon Q_{r})T_{Cr}\}\;
                                                          \nonumber \\
           & &+\int d^{2}\bar{\theta}\,
              \{m_{12r}^{2}(H_{1r}\epsilon H_{2r})^{\dagger}\bar{\theta}^{2}+
    m_{r}M_{r}^{\dagger}N_{r}^{\dagger}(1-B_{r}\bar{\theta}^{2})+
               h_{r}N_{r}^{\dagger}(H_{1r}\epsilon H_{2r})^{\dagger}+
               f_{Tr}(H_{2r}\epsilon Q_{r})^{\dagger}T_{Cr}^{\dagger}\}
                                                          \nonumber \\
           & &
\end{eqnarray}
where a subscript $r$ is used to distinguish the fields and couplings
defined at low energy from those defined at the cut-off $\Lambda$.
The non-trivial relations between these two sets of fields and couplings
are organized as follows:
\begin{eqnarray}
N_{r}&=&\sqrt{Z_{N}}N,\:   H_{2r}=\sqrt{Z_{H_{2}}}H_{2},\:
Q_{r}=\sqrt{Z_{Q}}Q,\:T_{Cr}=\sqrt{Z_{T_{C}}}T_{C} \\
m_{Nr}^{2}&=&-(m_{1}^{2}+m_{2}^{2}) \nonumber \\
m_{2r}^{2}&=&\frac{m_{2}^{2}-2\Sigma_{H_{2}}m_{T}^{2}}{1+\Sigma_{H_{2}}}
                                                      \nonumber \\
m_{12r}^{2}&=&\frac{m_{12}^{2}}{\sqrt{Z_{H_{2}}}}
                                                      \nonumber \\
m_{Qr}^{2}&=&\frac{(1-\Sigma_{Q})m_{T}^{2}-\Sigma_{Q}m_{2}^{2}}{1+\Sigma_{Q}}
                                                       \nonumber \\
m_{T_{Cr}}^{2}&=&
\frac{(1-\Sigma_{T_{C}})m_{T}^{2}-\Sigma_{T_{C}}m_{2}^{2}}{1+\Sigma_{T_{C}}}
                                                                  \\
f_{Tr}&=&\frac{f_{T}}{\sqrt{Z_{H_{2}}Z_{Q}Z_{T_{C}}}}             \\
m_{r}&=&\frac{1}{\sqrt{G}}\cdot
\frac{1}{\sqrt{\frac{N_{W}}{16\pi^{2}}\ln (\frac{\Lambda^{2}}{\mu_{E}^{2}})}}
                                                                   \\
h_{r}&=&\frac{1}{\sqrt{1+\Sigma_{H_{2}}}}\cdot
\frac{1}{\sqrt{\frac{N_{W}}{16\pi^{2}}\ln (\frac{\Lambda^{2}}{\mu_{E}^{2}})}}
\end{eqnarray}
Those fields and couplings that do not get renormalized are not listed.
Therefore, at low energy, the MSMHHC becomes the Non-Minimal SSM with the
couplings defined above. This Non-Minimal SSM will be called the (M+2)SSM.
Notice that these low-energy couplings depend only
on $G$, not on $h$ or $m$. Relations (10)$-$(13) will be useful later.

    From (13), the physical meaning of the low-energy Higgs self-coupling
$h_{r}$ is clear: it simply reflects the cut-off dependence of
$\cal L$$_{eff}$. In the limit $\mu_{E}\rightarrow\Lambda$,
$Z_{N}\rightarrow 0$ and the behavior of triviality is reproduced:
$h_{r}\rightarrow\infty$. Therefore, the issue of triviality is resolved
in the sense that the problem of triviality seems to arise only
because the composite $N$ is mistaken for a fundamental dynamical field
near the cut-off $\Lambda$. Of course, the complete resolution of triviality
would require us to go beyond the cut-off $\Lambda$ and to search for the
correct renormalizable theory at higher energy from which the MSMHHC can be
derived. However, this is beyond the scope of this paper.

    In (10), $m_{Nr}^{2}=\,-(m_{1}^{2}+m_{2}^{2})$ is negative. The
negative value of $m_{Nr}^{2}$, which is induced purely by the soft
SUSY-breaking scalar squared masses, suggests that condensation
should occur and $\phi_{N_{r}}$, the scalar component of $N_{r}$,
should develop a non-trivial VEV. This is consistent with the
observation made in \cite{Clark90,Carena92,Buchmuller84,Buchmuller82}
that, in the SUSY NJL model, there is no condensation in the supersymmetric
limit and soft SUSY-breaking scalar-mass terms must be included to trigger
condensation. (Details can be found in Section 4.) Therefore, the VEV
$<\phi_{N_{r}}>$ is induced only through supersymmetry breaking effects
and is naturally of order $F_{SUSY}$, the strength of
soft SUSY-breaking scalar mass. This is exactly the solution to the
naturalness problem of $\mu$, where $\mu$ is effectively generated by
$\mu_{r}=h_{r}<\phi_{N_{r}}>$ here. ($h_{r}$ is also natural since
its dependence on $\Lambda$ is logarithmic.)
Aside from the above argument, the
computations of $\mu_{r}$ versus $\Lambda$ and $\mu_{r}$ versus $F_{SUSY}$
are done in Sections 7$-$8, which constitute the concrete answer to the
\mbox{$\mu$ problem}.

    In order to simplify the computations but keep track of the same
physics at the same time, another simplification will be made in the
following sections: the SU(2) symmetry is turned off and all the
superfields are then singlets. That is, the MSMHHC of U(1) will be studied
instead of the MSMHHC of SU(2)$\times$U(1). Notice that the results
of this section, (2)$-$(13), are still valid under this simplification
except for $N_{W}$=1. Although the computations and figures presented in the
following sections are obtained by assuming the MSMHHC of U(1), their
physical features are shared by both the MSMHHC of U(1) and the MSMHHC of
SU(2)$\times$U(1). As we shall see later, the qualitative features
of the MSMHHC obtained under this simplification in this paper
remain true even in the general SU(2)$\times$U(1) case.
\section{The Effective Potential}
\hspace{0.8cm}In order to study the phenomena of condensation, one
must go beyond the tree-level computations of the SUSY NJL model.
Therefore, the effective potential of the MSMHHC will be computed
up to one loop, i.e., \mbox{$V_{eff}=V_{tree}+V_{1-loop}$}. Notice
that we are now working with the MSMHHC of U(1) symmetry. Using
the Coleman-Weinberg one-loop effective potential \cite{Coleman73}:
\begin{eqnarray}
V_{1-loop}&=&\frac{1}{2}STr\int\frac{d^{4}p}{(2\pi)^{4}}
             \ln (p^{2}+\hat{M}^{2})
\end{eqnarray}
where $STr(\hat{M}^{2})=Tr(\hat{M}_{B}^{2})-2\,Tr(\hat{M}_{F}^{2})$.
The convention of \cite{Falck86} will be adopted in the computations
of the spin-0 and spin-$\frac{1}{2}$ mass matrices, $\hat{M}_{B}$
and $\hat{M}_{F}$. Care should be taken in treating the superfield $N$
since it is static at tree level. The convention of component
notations for the chiral superfield $N$ is
\mbox{$N=\phi_{N}\,+\,\sqrt{2}\theta\psi_{N}\,+\,\theta\theta F_{N}$},
and it applies to other chiral superfields.

    Notice that $m_{12}^{2}(H_{1}\epsilon H_{2})\theta^{2}$ in (2)
can be written as $-\frac{m_{12}^{2}}{\sqrt{G}}M\theta^{2}$ with a
shift in $N$. $V_{tree}$ of the MSMHHC of U(1) symmetry can be computed
from $\cal L$$_{\Lambda}$ in (2) easily
\begin{eqnarray}
 V_{tree}&=&(h^{2}|\phi_{N}|^{2}+m_{1}^{2})|\phi_{H_{1}}|^{2}+
            (h^{2}|\phi_{N}|^{2}+m_{2}^{2})|\phi_{H_{2}}|^{2}+
            m_{M}^{2}|\phi_{M}|^{2}+m^{2}|\phi_{N}|^{2}
                                                     \nonumber \\
         & &+(m_{T}^{2}+f_{T}^{2}|\phi_{H_{2}}|^{2})
            (|\phi_{Q}|^{2}+|\phi_{T_{C}}|^{2})+
            f_{T}^{2}|\phi_{Q}|^{2}|\phi_{T_{C}}|^{2}
                                                     \nonumber \\
         & &+(h\phi_{N}\phi_{H_{1}})
            (f_{T}\phi_{Q}^{\dagger}\phi_{T_{C}}^{\dagger})+
            (h\phi_{N}^{\dagger}\phi_{H_{1}}^{\dagger})
            (f_{T}\phi_{Q}\phi_{T_{C}})
                                                    \nonumber \\
         & &+mB\phi_{M}\phi_{N}+
            mB\phi_{M}^{\dagger}\phi_{N}^{\dagger}+
            \frac{m_{12}^{2}}{\sqrt{G}}\phi_{M}+
            \frac{m_{12}^{2}}{\sqrt{G}}\phi_{M}^{\dagger}
                                                    \nonumber \\
         & &-F_{N}(m\phi_{M}+h\phi_{H_{1}}\phi_{H_{2}})
            -F_{N}^{\dagger}(m\phi_{M}^{\dagger}+
                  h\phi_{H_{1}}^{\dagger}\phi_{H_{2}}^{\dagger})
\end{eqnarray}
According to \cite{Falck86}, we then compute the field-dependent
squared-mass matrices, $\hat{M}_{B}^{2}$ and $\hat{M}_{F}^{2}$,
and the results are summarized as follows:

\lefteqn{\hat{M}_{F}^{2}=}
\[  \left( \begin{array}{cccc}
h^{2}|\phi_{N}|^{2} & 0 & (h\phi_{N}^{\dagger})
(f_{T}\phi_{T_{C}}) & (h\phi_{N}^{\dagger})(f_{T}\phi_{Q}) \\[.25in]
0 &
\begin{array}{l}
h^{2}|\phi_{N}|^{2} \\[-0.0in] +f_{T}^{2}(|\phi_{Q}|^{2}
+|\phi_{T_{C}}|^{2})
\end{array}
&
f_{T}^{2}\phi_{H_{2}}\phi_{Q}^{\dagger} &
f_{T}^{2}\phi_{H_{2}}\phi_{T_{C}}^{\dagger} \\[.25in]
(h\phi_{N})(f_{T}\phi_{T_{C}}^{\dagger}) &
f_{T}^{2}\phi_{H_{2}}^{\dagger}\phi_{Q} &
f_{T}^{2}(|\phi_{H_{2}}|^{2}+|\phi_{T_{C}}|^{2}) &
f_{T}^{2}\phi_{Q}\phi_{T_{C}}^{\dagger} \\[.25in]
(h\phi_{N})(f_{T}\phi_{Q}^{\dagger}) &
f_{T}^{2}\phi_{H_{2}}^{\dagger}\phi_{T_{C}} &
f_{T}^{2}\phi_{Q}^{\dagger}\phi_{T_{C}} &
f_{T}^{2}(|\phi_{H_{2}}|^{2}+|\phi_{Q}|^{2})
\end{array}  \right) \hspace{0.5in} (16) \]
\addtocounter{equation}{1}%

\vspace{0.3in}
\[ \hat{M}_{B}^{2} = \left( \begin{array}{cc}
 \hat{A} & \hat{B} \\
\hat{B}^{\dagger} & \hat{C}
\end{array} \right) \hspace{4.0in} (17) \]
\addtocounter{equation}{1}%

\vspace{0.3in}
\lefteqn{\hat{A}=}
\[  \left( \begin{array}{cccc}
h^{2}|\phi_{N}|^{2}+m_{1}^{2} &
-hF_{N}^{\dagger} &
0                 &
0                 \\[.25in]
-hF_{N}           &
\begin{array}{l}
h^{2}|\phi_{N}|^{2}+m_{2}^{2} \\[-0.0in]
+f_{T}^{2}(|\phi_{Q}|^{2}+|\phi_{T_{C}}|^{2})
\end{array}       &
0                 &
0                 \\[.25in]
0                 &
0                 &
\begin{array}{l}
h^{2}|\phi_{N}|^{2}+m_{2}^{2} \\[-0.0in]
+f_{T}^{2}(|\phi_{Q}|^{2}+|\phi_{T_{C}}|^{2})
\end{array}       &
-hF_{N}^{\dagger} \\[.25in]
0                 &
0                 &
-hF_{N}           &
h^{2}|\phi_{N}|^{2}+m_{1}^{2}
\end{array}  \right) \hspace{0.5in} (18) \]
\addtocounter{equation}{1}%

\vspace{0.3in}
\lefteqn{\hat{B}=}
\[  \left( \begin{array}{cccc}
(h\phi_{N}^{\dagger})(f_{T}\phi_{T_{C}})   &
0                                          &
(h\phi_{N}^{\dagger})(f_{T}\phi_{Q})       &
0                                          \\[.25in]
f_{T}^{2}\phi_{H_{2}}^{\dagger}\phi_{Q}^{\dagger} &
f_{T}^{2}\phi_{H_{2}}^{\dagger}\phi_{T_{C}}       &
f_{T}^{2}\phi_{H_{2}}^{\dagger}\phi_{T_{C}}^{\dagger} &
f_{T}^{2}\phi_{H_{2}}^{\dagger}\phi_{Q}    \\[.25in]
f_{T}^{2}\phi_{H_{2}}\phi_{Q}^{\dagger} &
f_{T}^{2}\phi_{H_{2}}\phi_{T_{C}}       &
f_{T}^{2}\phi_{H_{2}}\phi_{T_{C}}^{\dagger} &
f_{T}^{2}\phi_{H_{2}}\phi_{Q}           \\[.25in]
0                                       &
(h\phi_{N})(f_{T}\phi_{Q}^{\dagger})    &
0                                       &
(h\phi_{N})(f_{T}\phi_{T_{C}}^{\dagger})
\end{array}  \right) \hspace{0.5in} (19) \]
\addtocounter{equation}{1}%

\vspace{0.3in}
\lefteqn{\hat{C}=}
\[  \left( \begin{array}{cccc}
\begin{array}{l}
f_{T}^{2}(|\phi_{H_{2}}|^{2}+|\phi_{T_{C}}|^{2}) \\[-0.0in]
+m_{T}^{2}
\end{array}                              &
\begin{array}{l}
(h\phi_{N})(f_{T}\phi_{H_{1}}) \\[-0.0in]
+f_{T}^{2}\phi_{Q}\phi_{T_{C}}
\end{array}                              &
f_{T}^{2}\phi_{Q}\phi_{T_{C}}^{\dagger}  &
0                                        \\[.25in]
\begin{array}{l}
(h\phi_{N}^{\dagger})(f_{T}\phi_{H_{1}}^{\dagger}) \\[-0.0in]
+f_{T}^{2}\phi_{Q}^{\dagger}\phi_{T_{C}}^{\dagger}
\end{array}                              &
\begin{array}{l}
f_{T}^{2}(|\phi_{H_{2}}|^{2}+|\phi_{Q}|^{2}) \\[-0.0in] +m_{T}^{2}
\end{array}                              &
0                                        &
f_{T}^{2}\phi_{Q}\phi_{T_{C}}^{\dagger}  \\[.25in]
f_{T}^{2}\phi_{Q}^{\dagger}\phi_{T_{C}}  &
0                                        &
\begin{array}{l}
f_{T}^{2}(|\phi_{H_{2}}|^{2}+|\phi_{Q}|^{2}) \\[-0.0in] +m_{T}^{2}
\end{array}                              &
\begin{array}{l}
(h\phi_{N})(f_{T}\phi_{H_{1}}) \\[-0.0in]
+f_{T}^{2}\phi_{Q}\phi_{T_{C}}
\end{array}                              \\[.25in]
0                                        &
f_{T}^{2}\phi_{Q}^{\dagger}\phi_{T_{C}}  &
\begin{array}{l}
(h\phi_{N}^{\dagger})(f_{T}\phi_{H_{1}}^{\dagger}) \\[-0.0in]
+f_{T}^{2}\phi_{Q}^{\dagger}\phi_{T_{C}}^{\dagger}
\end{array}                              &
\begin{array}{l}
f_{T}^{2}(|\phi_{H_{2}}|^{2}+|\phi_{T_{C}}|^{2}) \\[-0.0in]
+m_{T}^{2}
\end{array}
\end{array}  \right)  \]
\addtocounter{equation}{1}%
\hspace{6.3in} (20)

    (14) can be integrated to give the following:
\begin{eqnarray}
V_{1-loop}&=&\frac{1}{32\pi^{2}}\{\; \Lambda^{2}STr(\hat{M}^{2})
           +\frac{1}{2}STr(\hat{M}^{4}\ln (\frac{\hat{M}^{2}}{\Lambda^{2}}))
             -\frac{1}{4}STr(\hat{M}^{4})
             +O(\frac{\hat{M}^{6}}{\Lambda^{2}}) \;\} \hspace{0.2in} (21)
\end{eqnarray}
In general, $\hat{M}^{2}$ is of order $F_{SUSY}^{2}$, and therefore
it's reasonable to neglect the last term in (21) because $F_{SUSY}\ll\Lambda$
in reality. The first term in (21) is also neglected because
$STr(\hat{M}^{2})$ depends only on the soft SUSY-breaking parameters, which
is just a constant contribution to $V_{eff}$. Denote the eigenvalues of
$\hat{M}_{B}^{2}$ and $\hat{M}_{F}^{2}$ as $\omega_{Bi}^{2}$ and
$\omega_{Fi}^{2}$ respectively, and $V_{1-loop}$ can be written as follows:
\begin{eqnarray}
V_{1-loop}&=&\frac{1}{32\pi^{2}}\sum_{Bi=1}^{8} \{\;
         \frac{1}{2}\omega_{Bi}^{4}\ln (\frac{\omega_{Bi}^{2}}{\Lambda^{2}})
            -\frac{1}{4}\omega_{Bi}^{4} \;\} \nonumber \\
          & &-\frac{1}{16\pi^{2}}\sum_{Fi=1}^{4} \{\;
         \frac{1}{2}\omega_{Fi}^{4}\ln (\frac{\omega_{Fi}^{2}}{\Lambda^{2}})
            -\frac{1}{4}\omega_{Fi}^{4} \;\}
\end{eqnarray}
In general, $\hat{M}_{B}^{2}$ and $\hat{M}_{F}^{2}$ are non-trivial
large matrices, and therefore numerical methods are needed in
order to compute $\omega_{Bi}^{2}$ and $\omega_{Fi}^{2}$. The above
results will be useful later. In fact, our computations of $V_{eff}$
will be verified in \mbox{Section 4} by comparing a special case of
ours with that of \cite{Binetruy94}. The extension of the computations
(15)$-$(22) to the MSMHHC of SU(2)$\times$U(1) has been worked out,
too. However, this extension is straightforward, and therefore it is
not given here.
\section{The Vacuum of the MSMHHC}
\hspace{0.8cm}Due to phenomenological considerations, we are interested
in the vacuum configuration of the MSMHHC
where $\phi_{H_{1}}$, $\phi_{H_{2}}$,
$\phi_{M}$ and $\phi_{N}$ develop non-trivial VEV's, but the other scalar
fields do not. That is, the SU(2)$\times$U(1)$\;\rightarrow\;$U(1)
electroweak symmetry breaking accompanies the Higgs-Higgs condensation
for the MSMHHC of SU(2)$\times$U(1) symmetry. For the MSMHHC of U(1) symmetry,
this vacuum configuration means that the spontaneous breaking of U(1)
accompanies the Higgs-Higgs condensation.
This MSMHHC vacuum is obtained by finding the extremum of $V_{eff}$. Thanks
to $<\phi_{Q}>=<\phi_{T_{C}}>=0$, the extremization of $V_{eff}$ can be
performed exactly, and it leads to a set of five vacuum
constraints. The imaginary parts of these vacuum constraints
are used to fix the relative phases among different VEV's.
The above computations have been performed for the MSMHHC of U(1)
symmetry and the results are as follows. In this paper, we always
take $m$, $h$, $B$ and $f_{T}$ to be positive without loss of generality.
\begin{eqnarray}
<\phi_{H_{1}}>&=&v_{1},\:\: <\phi_{H_{2}}>=v_{2}e^{i\varphi},\:\:
<\phi_{M}>=\,-v_{M}e^{i\varphi},
                                                      \nonumber \\
<\phi_{N}>&=&\,v_{N}e^{-i\varphi},\:\: <F_{N}>=\,v_{F_{N}}e^{-i\varphi},
                                                      \nonumber \\
<\phi_{Q}>&=&<\phi_{T_{C}}>=0
\end{eqnarray}
where $v_{1}$, $v_{2}$, $v_{M}$, $v_{N}$ and $v_{F_{N}}$ are positive.
{}From now on, (23) will be called the MSMHHC vacuum, and we always choose
the special case: $\varphi=0$. When $\varphi=0$, notice that
$<\phi_{M}>$ and $<\phi_{N}>$ are exactly out of phase.
With this MSMHHC vacuum,
there are still five real vacuum constraints, (24)$-$(28), coming from
the variation of $V_{eff}$ (the effective potential of the MSMHHC
of U(1), (15)$-$(22)) with respect to $v_{1}$, $v_{2}$, $v_{M}$,
$v_{N}$ and $v_{F_{N}}$ respectively.
\begin{eqnarray}
hv_{F_{N}}v_{2}&=&(h^{2}v_{N}^{2}+m_{1}^{2})v_{1}
\,-\,\frac{f_{T}}{32\pi^{2}}(hv_{N})
\{\, \omega_{3}^{2}\ln (\frac{\Lambda^{2}}{\omega_{3}^{2}})
\,-\, \omega_{4}^{2}\ln (\frac{\Lambda^{2}}{\omega_{4}^{2}}) \,\}
                                                               \\
hv_{F_{N}}v_{1}&=&(h^{2}v_{N}^{2}+m_{2}^{2})v_{2}
\,-\,\frac{f_{T}^{2}v_{2}}{16\pi^{2}}
\{\, \omega_{3}^{2}\ln (\frac{\Lambda^{2}}{\omega_{3}^{2}})
\,+\, \omega_{4}^{2}\ln (\frac{\Lambda^{2}}{\omega_{4}^{2}}) \nonumber \\
& &\hspace{2.5in} \,-\, 2\omega_{6}^{2}
   \ln (\frac{\Lambda^{2}}{\omega_{6}^{2}}) \,\}
                                                                \\
hv_{F_{N}}&=&\,-\sqrt{G}m_{M}^{2}v_{M}+B(hv_{N})+m_{12}^{2}
\end{eqnarray}
\begin{eqnarray}
-\,\frac{B}{\sqrt{G}}v_{M}
&+&hv_{N}(v_{1}^{2}+v_{2}^{2}+\frac{1}{G})
\,-\,\frac{hv_{N}}{16\pi^{2}}\{\, \omega_{1}^{2}
\ln (\frac{\Lambda^{2}}{\omega_{1}^{2}})
\,+\, \omega_{2}^{2}\ln (\frac{\Lambda^{2}}{\omega_{2}^{2}}) \nonumber \\
&-&2\omega_{5}^{2}\ln (\frac{\Lambda^{2}}{\omega_{5}^{2}})
\,+\, (\frac{f_{T}v_{1}}{2hv_{N}})\omega_{3}^{2}
\ln (\frac{\Lambda^{2}}{\omega_{3}^{2}})
\,-\, (\frac{f_{T}v_{1}}{2hv_{N}})\omega_{4}^{2}
\ln (\frac{\Lambda^{2}}{\omega_{4}^{2}}) \,\}\,=0
\hspace{0.5in} (27) \nonumber
\end{eqnarray}
\addtocounter{equation}{1}%
\begin{eqnarray}
v_{1}v_{2}&-&\frac{1}{\sqrt{G}}v_{M}
\,+\,\frac{1}{32\pi^{2}}
(\frac{hv_{F_{N}}}
{\sqrt{\frac{1}{4}(m_{1}^{2}-m_{2}^{2})^{2}+h^{2}v_{F_{N}}^{2}}})
\{\, \omega_{1}^{2}\ln (\frac{\Lambda^{2}}{\omega_{1}^{2}})
\,-\, \omega_{2}^{2}\ln (\frac{\Lambda^{2}}{\omega_{2}^{2}}) \,\}\,=0
\hspace{0.5in} (28) \nonumber
\end{eqnarray}
\addtocounter{equation}{1}%
where
\begin{eqnarray}
\omega_{1}^{2}&=&h^{2}v_{N}^{2}+\frac{1}{2}(m_{1}^{2}+m_{2}^{2})
+\sqrt{\frac{1}{4}(m_{1}^{2}-m_{2}^{2})^{2}+h^{2}v_{F_{N}}^{2}} \\
\omega_{2}^{2}&=&h^{2}v_{N}^{2}+\frac{1}{2}(m_{1}^{2}+m_{2}^{2})
-\sqrt{\frac{1}{4}(m_{1}^{2}-m_{2}^{2})^{2}+h^{2}v_{F_{N}}^{2}} \\
\omega_{3}^{2}&=&f_{T}^{2}v_{2}^{2}+m_{T}^{2}+(hv_{N})(f_{T}v_{1}) \\
\omega_{4}^{2}&=&f_{T}^{2}v_{2}^{2}+m_{T}^{2}-(hv_{N})(f_{T}v_{1}) \\
\omega_{5}^{2}&=&h^{2}v_{N}^{2} \\
\omega_{6}^{2}&=&f_{T}^{2}v_{2}^{2}
\end{eqnarray}
(24)$-$(34) uniquely define the study of the MSMHHC vacuum
for the MSMHHC of U(1) symmetry.
The eigenvalues (of multiplicity 2) of $\hat{M}_{B}^{2}$
and $\hat{M}_{F}^{2}$ evaluated at the MSMHHC vacuum
are denoted here as ($\omega_{1}^{2},\ldots ,\omega_{4}^{2}$) and
($\omega_{5}^{2},\: \omega_{6}^{2}$) respectively. Numerical
methods are needed in order to study (24)$-$(34).

    In comparison with the usual SUSY NJL model (for example,
the MTCSSM \cite{Binetruy94}), the MSMHHC has two more vacuum constraints,
i.e., the constraints (24) and (25), which corresponds to the
fact that the constituent fields ($H_{1}$ and $H_{2}$) of
the condensates also develop non-trivial VEV's ($v_{1}$ and $v_{2}$)
in the formulation of the MSMHHC. In the MTCSSM, the constituent fields of
the condensates are the top superfields, which do not develop
non-trivial VEV's. Therefore, when \mbox{$v_{1}=v_{2}=0$}, constraints (24)
and (25) are washed away and we recover the usual results.
If $f_{T}=0$ and $B=m_{12}=0$ are further assumed, the constraints
(26)$-$(28) will be the same as those obtained in \cite{Binetruy94}
for the MTCSSM essentially, and $G$ will satisfy the usual SUSY
gap equation \cite{Binetruy94}.

     However, with $v_{1},\;v_{2}\neq 0$, the MSMHHC is indeed more
complicated than the usual SUSY NJL model. Due to the unique feature that
the MSMHHC has two more vacuum constraints, an immediate implication is
that the MSMHHC should be more predictive than the usual SUSY NJL model
(e.g., the MTCSSM). In fact, there is one
more physical constraint for the MSMHHC:
\mbox{$\sqrt{v_{1r}^{2}+v_{2r}^{2}}=\frac{250}{\sqrt{2}}$ GeV} at low energy.
Therefore, together with the constraints (24)$-$(28), the independent input
parameters required by the MSMHHC are:
\begin{equation}
\Lambda,\; f_{T}\:\:\mbox{and}\:\:\:
(\; m_{1},\; m_{2},\; m_{M},\; m_{T},\; m_{12},\; B \;)
\end{equation}
Once (35) is specified, everything is determined. Notice
that the predictions of the MSMHHC are essentially determined
by the soft SUSY-breaking terms only. With (35) specified, the
determination of the following quantities is always interesting:
\begin{equation}
G,\;\: v_{M},\;\: hv_{N},\;\: hv_{F_{N}},\;\: \tan\beta =\frac{v_{2}}{v_{1}}
\end{equation}
Unlike the MTCSSM, the four-Fermi coupling $G$ is determined from
(35) and is not a free parameter in MSMHHC. This suggests that,
within the framework of the MSMHHC, the origin of the four-field
interaction, $G(H_{1}\epsilon H_{2})^{\dagger}(H_{1}\epsilon H_{2})$,
should be closely related to the supersymmetry breaking. A systematic
numerical study of the MSMHHC vacuum throughout the parameter space
will be given in \mbox{Sections 6$-$9}.
\section{\mbox{The Mass Spectrum and the Lightest Higgs Boson}}
\hspace{0.8cm}To determine the full spectrum of the scalar particles,
we need to compute the full squared-mass matrix from $V_{eff}$,
that is, the second derivatives of $V_{eff}$ evaluated at the MSMHHC
vacuum. Due to the explicit dependence of $V_{eff}$ on the
auxiliary field $F_{N}$, the computation of the second derivative is
a little complicated:
\pagebreak
\begin{eqnarray}
\frac{d^{2}V_{eff}}{dAdB}&=&\frac{\partial^{2}V_{eff}}{\partial A\partial B}
 +\frac{\partial^{2}V_{eff}}{\partial A\partial F_{N}}
 (\frac{\partial F_{N}}{\partial B})
 +\frac{\partial^{2}V_{eff}}{\partial A\partial F_{N}^{\dagger}}
 (\frac{\partial F_{N}^{\dagger}}{\partial B})
                                              \nonumber \\
&&+\frac{\partial^{2}V_{eff}}{\partial B\partial F_{N}}
(\frac{\partial F_{N}}{\partial A})
+\frac{\partial^{2}V_{eff}}{\partial B\partial F_{N}^{\dagger}}
(\frac{\partial F_{N}^{\dagger}}{\partial A})
                                              \nonumber \\
&&+\frac{\partial^{2}V_{eff}}{\partial F_{N}^{2}}
             (\frac{\partial F_{N}}{\partial A})
             (\frac{\partial F_{N}}{\partial B})
+\frac{\partial^{2}V_{eff}}{\partial F_{N}^{\dagger^{2}}}
             (\frac{\partial F_{N}^{\dagger}}{\partial A})
             (\frac{\partial F_{N}^{\dagger}}{\partial B})
                                              \nonumber \\
&&+\frac{\partial^{2}V_{eff}}{\partial F_{N}\partial F_{N}^{\dagger}}
  [\, (\frac{\partial F_{N}}{\partial A})
  (\frac{\partial F_{N}^{\dagger}}{\partial B})
   \,+\,(\frac{\partial F_{N}^{\dagger}}{\partial A})
   (\frac{\partial F_{N}}{\partial B}) \,]
\end{eqnarray}
where $A$ and  $B$ represent the scalar fields $\phi_{H_{1}}$,
$\phi_{H_{2}}$, $\phi_{M}$, $\phi_{N}$, $\phi_{Q}$, $\phi_{T_{C}}$,
or their complex conjugates. The constraint
$\frac{\partial V_{eff}}{\partial F_{N}}=0$ has been used in deriving (37).
Terms like $\frac{\partial F_{N}}{\partial A}$
can be obtained by differentiating
$\frac{\partial V_{eff}}{\partial F_{N}}=0$.
However, (37) is useful only if eigenvalues of the large matrices
involved in $V_{eff}$ (22) can be solved exactly. In general, $V_{eff}$ and
its second derivatives have to be computed by numerical methods.
These numerical computations are time-consuming because,
given any scalar field configuration, the evaluation of $V_{eff}$ requires
solving a \mbox{2-dim} extremization problem numerically due to the
constraint $\frac{\partial V_{eff}}{\partial F_{N}}=0$.
More details about the computation of the mass spectrum will be
given in \mbox{Section 7}. As expected, the numerically computed
mass spectrum for the MSMHHC of U(1) contains exactly one massless
particle due to the broken U(1) symmetry. We have
also computed the mass spectrum for the MSMHHC of SU(2)$\times$U(1),
and there are three massless particles, i.e., three Goldstone bosons
which will be absorbed by gauge bosons if SU(2)$\times$U(1) is gauged.
The systematic numerical study of the mass spectrum for the
MSMHHC of U(1) symmetry throughout the parameter
space will be given in \mbox{Sections 7$-$9}.

    Because the mass of the lightest Higgs boson ($m_{LH}$) is one of our
main concerns, it's definitely important to ask how $m_{LH}$ depends on the
top Yukawa interaction. The best way to answer this question is to turn off
the top Yukawa interaction ($f_{T}=0$), and therefore the MSMHHC of U(1)
is reduced to a pure Higgs sector. To reveal a unique feature of the MSMHHC
of U(1) with $f_{T}=0$, we write down
(24), (25) and (30) with $f_{T}$ set to zero:
\begin{eqnarray}
hv_{F_{N}}v_{2}&=&(h^{2}v_{N}^{2}+m_{1}^{2})v_{1}
                                                \\
hv_{F_{N}}v_{1}&=&(h^{2}v_{N}^{2}+m_{2}^{2})v_{2}
                                                \\
\omega_{2}^{2}&=&h^{2}v_{N}^{2}+\frac{1}{2}(m_{1}^{2}+m_{2}^{2})
-\sqrt{\frac{1}{4}(m_{1}^{2}-m_{2}^{2})^{2}+h^{2}v_{F_{N}}^{2}}
\end{eqnarray}
(38) and (39) are the vacuum constraints corresponding to the
variations of $V_{eff}$ with respect to $v_{1}$ and $v_{2}$,
which do not exist in the usual SUSY NJL model. Notice that, with
$f_{T}=0$, (38) and (39) are actually the tree-level vacuum constraints, and
remain unaffected by one-loop corrections. $\omega_{2}^{2}$
is an eigenvalue of the tree-level squared-mass matrix $\hat{M}_{B}^{2}$.
Remember that we have been working with the MSMHHC of U(1), and therefore
$\omega_{2}^{2}$ in (40) has multiplicity 2. Notice that the validity of
(38)$-$(40) is not limited the MSMHHC of U(1). In fact, it is straightforward
to verify that (38)$-$(40) also hold for the MSMHHC of SU(2)$\times$U(1), and
$\omega_{2}^{2}$ has multiplicity 4 in the case of SU(2)$\times$U(1).
Combining (38) and (39) with (40), one has $\omega_{2}^{2}=0$. It means that,
at tree level, there are two massless particles for the MSMHHC of U(1)
(or, four massless particles for the MSMHHC of SU(2)$\times$U(1)). In either
case, there is one more massless particle than what is usually expected.
However, the above tree-level results are not conclusive, and we must
resort to one-loop contributions, $V_{eff}=V_{tree}+V_{1-loop}$.

    With $f_{T}=0$, $V_{eff}$ can be computed exactly, and (37) is useful in
computing the second derivatives of $V_{eff}$ at the vacuum. Because these
computations are trivial and lengthy, their details are not presented here.
The full spectrum is then obtained by solving the eigenvalues of the full
squared-mass matrix, which is composed of the second derivatives of
$V_{eff}$ evaluated at the vacuum. Our computations for the MSMHHC
of U(1) indicates that certain one-loop contributions cancel during the
computation, and there are still two massless particles. We have also
computed the case of SU(2)$\times$U(1), and there are
four massless particles. That is, besides the Goldstone bosons,
there is always one more massless particle for the MSMHHC of either U(1) or
SU(2)$\times$U(1), where this additional massless particle is
just the lightest Higgs. Therefore, the unique feature of the MSMHHC with
$f_{T}=0$ is: the lightest Higgs remains massless, at least to one-loop order.
Although we are not able to give a formal proof of the above result in
a few sentences, this unique feature is indeed well established by the
reliable computations based on (37) and (15)$-$(22).

    An obvious way to make the lightest Higgs massive is to violate
the vacuum constraints (38) and (39). If they are violated, then
$\omega_{2}^{2}=0$ will no longer hold, and we shall have exactly one
massless particle for the MSMHHC of U(1) (or, three massless particles
for the MSMHHC of SU(2)$\times$U(1)). Furthermore,
the lightest Higgs will receive a mass proportional to the amount of
violation if it is small. According to (24) and (25), turning on the top
Yukawa interaction does violate (38) and (39). Therefore, the violation of
$\omega_{2}^{2}=0$ due to $f_{T}$ is roughly proportional to
\mbox{$f_{T}^{2}(h^{2}v_{N}^{2}\sigma_{1}+2m_{T}^{2}\sigma_{2})$}, where
\begin{eqnarray}
\sigma_{1}&=&\frac{h^{2}v_{N}^{2}+m_{2}^{2}}
{h^{2}v_{N}^{2}+\frac{1}{2}(m_{1}^{2}+m_{2}^{2})}
                                        \nonumber \\
\sigma_{2}&=&\frac{h^{2}v_{N}^{2}+m_{1}^{2}}
{h^{2}v_{N}^{2}+\frac{1}{2}(m_{1}^{2}+m_{2}^{2})}
\end{eqnarray}
So, we have the following rough qualitative estimate for $m_{LH}$:
\begin{eqnarray}
m_{LH}^{2}\;\propto\;f_{T}^{2}(h^{2}v_{N}^{2}\sigma_{1}+2m_{T}^{2}\sigma_{2})
\end{eqnarray}
This qualitative estimate is good when the amount of violation
(i.e., $m_{LH}$) is not very large, and the above argument indicates
that (42) is shared by both the MSMHHC of U(1) and the MSMHHC of
SU(2)$\times$U(1). (42) will be shown
to be compatible with the results obtained in \mbox{Sections 8 and 9.}
(42) shows that $m_{LH}$ is proportional to the mass
of the top quark, which reveals an important feature of the MSMHHC: heavy
top quark always implies comparatively large $m_{LH}$.
This is really a nice feature from the viewpoint of the second issue.
(42) also indicates that $m_{LH}$ is proportional to $F_{SUSY}$
(the strength of soft SUSY breaking). As we shall see in \mbox{Section 8},
$m_{LH}$ is indeed proportional to $F_{SUSY}$, and saturates
when $F_{SUSY}$ becomes very large.
\section{The Parameter Space and Low-Energy \mbox{Quantities}}
\hspace{0.8cm}In \mbox{Sections 6$-$9}, we always work with the
MSMHHC of U(1). However, there is no loss of generality because
the qualitative features obtained in \mbox{Sections 6$-$9}
are shared by both the MSMHHC of U(1) and the MSMHHC of
SU(2)$\times$U(1). Therefore, the MSMHHC of U(1) symmetry won't be
emphasized unless it is necessary.
For convenience, we always specify the parameter space
at $\Lambda$ according to (35). That is, $\Lambda,\; f_{T}$ and
\mbox{$(m_{1},\, m_{2},\, m_{M},\, m_{T},\, m_{12},\, B)$}
define the full parameter space of the MSMHHC at the scale $\Lambda$,
where all the parameters are taken to be positive without loss of
generality. At any point of this parameter space, we first compute
the relevant physical quantities at $\Lambda$, and then renormalize
all the parameters and physical quantities down to low energy properly
according to \mbox{(9)$-$(13)}. The low-energy ones are
distinguished from the ones at $\Lambda$ by a subscript $r$,
and therefore care should be taken. A systematic study of the full
parameter space is not feasible unless reasonable physical
constraints are imposed in order to reduce the number of free
parameters. Next, we shall discuss the physical constraints assumed
here one by one.

    The first reasonable constraint is to require that all the low-energy
soft SUSY-breaking scalar squared masses be non-negative except for
the induced scalar squared mass $m_{N_{r}}^{2}$. When applied to (10),
it leads to two non-trivial constraints: \mbox{$m_{2r}^{2}\geq 0$} and
\mbox{$m_{T_{r}}^{2}\geq 0$}, where
\mbox{$\,m_{T_{r}}^{2}=m_{Q_{r}}^{2}=m_{T_{Cr}}^{2}\,$}
and \mbox{$\,\Sigma_{T}=\Sigma_{Q}=\Sigma_{T_{C}}=\Sigma_{H_{2}}\,$}
due to \mbox{$N_{W}=1$}.
These two constraints can be rewritten in terms of $\Sigma_{T}$:
\begin{eqnarray}
&&\Sigma_{T}\leq \frac{m_{2}^{2}}{2m_{T}^{2}}\:\:\: \mbox{and} \:\:\:
\Sigma_{T}\leq \frac{m_{T}^{2}}{m_{2}^{2}+m_{T}^{2}},
                                   \nonumber \\
&&\Sigma_{T}=\frac{f_{T}^{2}}{16\pi^{2}}\ln (\frac{\Lambda^{2}}{\mu_{E}^{2}})
\end{eqnarray}
where $\mu_{E}$ is chosen to be the low-energy scale, i.e.,
\mbox{$\mu_{E}\ll \Lambda$}. In the
case of either \mbox{$m_{2}^{2}\gg m_{T}^{2}$}
or \mbox{$m_{2}^{2}\ll m_{T}^{2}$}, (43) requires
\mbox{$\Sigma_{T}=\frac{f_{T}^{2}}{16\pi^{2}}\ln
(\frac{\Lambda^{2}}{\mu_{E}^{2}})\ll 1$}.
\mbox{$\frac{f_{T}^{2}}{16\pi^{2}}\ln (\frac{\Lambda^{2}}{\mu_{E}^{2}})\ll 1$}
is not favored by phenomenology because it may lead to small $f_{T}$,
and small $f_{T}$ contradicts with the present observation of heavy top quark.
The absolute upper bound on $\Sigma_{T}$ can be derived from (43):
\begin{eqnarray}
&&\Sigma_{T}=\frac{f_{T}^{2}}{16\pi^{2}}
\ln (\frac{\Lambda^{2}}{\mu_{E}^{2}})\;\leq\;\frac{1}{2},
\:\:\:\:\mbox{and}                \nonumber \\
&&\Sigma_{T}=\frac{f_{T}^{2}}{16\pi^{2}}
\ln (\frac{\Lambda^{2}}{\mu_{E}^{2}})\;=\;\frac{1}{2}
\:\:\:\mbox{when}\:\:\: m_{2}^{2}=m_{T}^{2}.
\end{eqnarray}
Due to the fact that top quark is heavy, the first constraint requires us to
choose $m_{2}^{2}\approx m_{T}^{2}$ naturally. In practice, it is found that
\mbox{$\frac{1}{2}<\frac{m_{2}}{m_{T}}<2$} is required when the
physical \mbox{top-quark mass $m_{top}=180$ GeV}. Therefore,
\mbox{$m_{2}^{2}=m_{T}^{2}$} is chosen in the study of the parameter space.

    (44) leads to another important result when it is combined with (13)
\mbox{($\Sigma_{H_{2}}=\Sigma_{T}$)}:
\begin{eqnarray}
h_{r}&=&\frac{1}{\sqrt{1+\Sigma_{T}}}\cdot
  \frac{1}{\sqrt{\frac{1}{16\pi^{2}}\ln (\frac{\Lambda^{2}}{\mu_{E}^{2}})}}
\;\geq\;\frac{2}{\sqrt{3}}f_{T}
\end{eqnarray}
(45) implies that heavy top quark (i.e., large $f_{T}$) always leads to
strongly-interacting low-energy Higgs sector (i.e., large $h_{r}$), a unique
feature of MSMHHC. Since a strongly-interacting low-energy Higgs sector always
leads to large $m_{LH}$, (45) actually re-establishes the
conclusion obtained in \mbox{Section 5}.

    The second constraint is to fix the physical top-quark mass $m_{top}$ as
\mbox{164 GeV $\sim$ 180 GeV}, and therefore $f_{T}$ is fixed in the study
of the parameter space. Generally, it is found that $1< f_{T}<2$.
Therefore, (44) can be interpreted in a different way by rewriting it
as follows:
\begin{eqnarray}
&&\Lambda\;\leq\;\mu_{E}\cdot e^{\frac{4\pi^{2}}{f_{T}^{2}}}
\end{eqnarray}
(46) leads to an upper bound on $\Lambda$ and, due to the second
constraint \mbox{($1< f_{T}<2$)}, it implies that the MSMHHC
should be an effective intermediate-scale model, i.e.,
\mbox{weak scale $\ll\,\Lambda\,\ll$ Planck scale}. Finally, we
shall fix $m_{M}^{2}$ by $m_{M}^{2}=m_{1}^{2}+m_{2}^{2}$. According
to (4), $M$ is the composite of $H_{1}$ and $H_{2}$. Therefore, it
is reasonable to assume that the SUSY breaking felt by $M$ is the
sum of those felt by $H_{1}$ and $H_{2}$ \cite{Clark90,Carena92}.

    However, there is a fine-tuning problem associated with the soft
SUSY-breaking parameters. One way to express this fine-tuning problem
is to fix all the parameters except for $B$. \mbox{(24)$-$(28)} are then
solved numerically with respect to different choices of $B$, and
the results indicate that $B$ must satisfy the following inequality:
\begin{eqnarray}
&&B_{lower}\;\leq\;B\;<\;B_{upper}
\end{eqnarray}
where $B_{upper}$ and $B_{lower}$ depend on all the parameters except for $B$.
In Fig.2, (47) is plotted with respect to different
choices of $F_{SUSY}$, where the other parameters are chosen as:
\mbox{$\Lambda=2.5\times 10^{4}$ TeV}, \mbox{$f_{T}=1.6\,$},
\mbox{$m_{1}=2\,F_{SUSY}$}, \mbox{$m_{2}=m_{T}=F_{SUSY}$},
\mbox{$m_{M}=\sqrt{5}\,F_{SUSY}$}, \mbox{$m_{12}=\frac{1}{2}\,F_{SUSY}$}.
$B\geq B_{lower}$ is a gap equation, and therefore this lower bound is
natural. That is, in studying (24)$-$(28), \mbox{$B<B_{lower}$}
results in no solution. (24)$-$(28) will result in exactly one solution for
the vacuum if \mbox{$B=B_{lower}$}, and two solutions if
\mbox{$B>B_{lower}$}. However, by checking the mass spectrum, the vacuum will
develop instability if $B>B_{upper}$.
In general, $\frac{B_{upper}-B_{lower}}{B}\approx \frac{5}{100}$.
Therefore, the upper bound $B_{upper}$ is unnatural and $B$ needs fine tuning.
This fine-tuning problem arises mainly due to the fact that the
MSMHHC has two more vacuum constraints (24) and (25).
Since this fine tuning is not severe, it suggests that a better choice of
non-renormalizable interactions may be able to relax this fine tuning
by modifying the vacuum constraints (24)$-$(28) properly.
The freedom of choosing non-renormalizable interactions
without violating symmetries has been an objection to the model of
top condensation because this arbitrariness
undermines its predictive power \cite{Lindner93,Suzuki90}.
The same objection applies to MSMHHC. However, by the requirement of
no fine tuning, the possible choices of
non-renormalizable interactions can be highly reduced,
and therefore this objection
may be partially resolved. From this viewpoint,
this fine-tuning problem is a useful
guide to the future model-building of the MSMHHC rather than an obstacle.
For the present, we simply choose $B=B_{lower}$ in the study of
the parameter space. According to this choice, $m_{LH}$ computed in
\mbox{Sections 7$-$9} is actually the upper bound.

    According to the above constraints and choices, the parameter space
under study contains only four free parameters:
\begin{eqnarray}
&&\Lambda\:\:\: \mbox{and}\:\:\: (\;m_{1},\; m_{2},\; m_{12}\;)
\end{eqnarray}
where the other parameters are determined by:
\begin{eqnarray}
&&m_{T}=m_{2},\:\:  m_{M}=\sqrt{m_{1}^{2}+m_{2}^{2}},\:\: B=B_{lower}
\end{eqnarray}
and $f_{T}$ is determined by the requirement that the physical top-quark
mass $m_{top}$ be \mbox{164 GeV $\sim$ 180 GeV}. Based on (48) and (49), a
systematic study of the parameter space will be given in
\mbox{Sections 7$-$9}.

    We have described in Sections 3$-$6 how to compute the physics of
the MSMHHC using the effective potential. However, the approach using RGE's
is adopted in many studies of the top-condensate models
\cite{Clark90,Carena92}. Therefore, it is worth mentioning the RGE approach
to the MSMHHC and comparing it with the present approach. It has been shown
that the low-energy effective theory of the MSMHHC is the (M+2)SSM, whose
details are given in (8)$-$(13). The spirit of the RGE approach is to work
with the low-energy theory (M+2)SSM instead of the MSMHHC, and to translate
the high-energy features of MSMHHC into the high-energy boundary condition
of the (M+2)SSM RGE's. That is, all the RGE approach needs is just the
(M+2)SSM and its boundary condition at the cut-off $\Lambda$. According to
(12) and (13), it is obvious that the correct boundary condition (the
compositeness condition) for the (M+2)SSM is:
\begin{equation}
  h_{r}=\infty,\;\; m_{r}=\infty\;\; \mbox{and}\;\;
  \frac{h_{r}}{m_{r}}=\mbox{finite}\;\;\;\mbox{at}\;\;\mu_{E}=\Lambda.
\end{equation}
As discussed in Section 2, this compositeness condition can be viewed as
the triviality problem of the (M+2)SSM, which indicates the existence of
composite fields in the (M+2)SSM. Besides, the $h_{r}$ computed from the
compositeness condition is just the so-called "triviality bound"
\cite{yywu,Moroi92}. Therefore, through the RGE approach, the study of
triviality bounds merges in the study of condensate models nicely.
As for technical aspects, it's easier to implement the RGE approach than
the effective potential approach, especially when the full content of the
standard model is included. These viewpoints do make the (M+2)SSM an
interesting low-energy model. We plan to publish the study of the RGE
approach to the MSMHHC (including gauge fields) and the study of triviality
bounds in the (M+2)SSM in a separate paper.

     However, it's necessary to explain why the RGE approach is not adopted
in the present paper in order to emphasize the physical significance of
the effective potential approach. We may start with the question: "Can the
RGE approach faithfully embrace all the physical features of the high-energy
condensate model in a natural way?" Since there is no one-to-one
correspondence between the low-energy effective models and the high-energy
models, the answer is no because the RGE approach is based only on the
low-energy effective model with the compositeness condition, and the
compositeness condition alone is not enough to embrace all the physical
features of the high-energy condensate model. For example, the SUSY gap
equation, which is essential in our solution to the \mbox{$\mu$ problem},
does not come out of the RGE approach naturally. On the other hand, the
effective potential approach is the natural and faithful way to express the
high-energy physics. Since the Higgs-Higgs condensation is a new idea, we
choose the effective potential approach in the present paper in order to
give a complete presentation of the physics.
\section{The Dependence of Low-Energy \mbox{Quantities}
         on \mbox{the Cut-off $\Lambda$}}
\hspace{0.8cm}The general computational procedure
has been described in detail, and its basic steps are summarized
as follows. Given a point of the parameter space according to
(48) and (49), the first step is to solve (24)$-$(28) for the MSMHHC
vacuum numerically, and therefore all the non-trivial VEV's, $G$ and
$B_{lower}$ are determined. Next, according to \mbox{Sections 3 and 5},
the full mass spectrum is computed. Finally, according to (9)$-$(13),
all the physical quantities and parameters are renormalized down to
the low-energy scale $\mu_{E}=v_{N_{r}}$, where $v_{N_{r}}$ is of
order the physical mass of the lightest Higgs $m_{LH}$. We always require
\mbox{$m_{top}$=164 or 180 GeV.} ($m_{top}$: the physical mass of top quark),
and therefore the choice of $f_{T}$ should be consistent with this $m_{top}$.

    The dependence on $\Lambda$ is illustrated with a typical example,
where its parameters are chosen as: \mbox{$F_{SUSY}=800$ GeV},
\mbox{$m_{1}=2\,F_{SUSY}$}, \mbox{$m_{2}=F_{SUSY}$},
\mbox{$m_{12}=\frac{1}{2}\,F_{SUSY}$},
\mbox{$m_{top}$=164 or 180 GeV}, and $\Lambda$ is left free.
The physical mass of the lightest Higgs
boson $m_{LH}$ versus the cut-off $\Lambda$ is plotted in Fig.3.
With \mbox{$10^{4}\,\mbox{GeV}\,<\,\Lambda\,<\,10^{8}\,\mbox{GeV}$},
the mass of the lightest Higgs can be as large as
\mbox{$150\,\mbox{GeV}\,<\,m_{LH}\,<\,400\,\mbox{GeV}$}
in Fig.3. Compared to the MSSM or the NMSSM, the MSMHHC is
indeed able to predict much larger $m_{LH}$. Therefore, the MSMHHC widens
the Higgs search, and should be taken more seriously if the Higgs signal
is still absent in the near future. In general,
$m_{LH}$ decreases as $\Lambda$ increases. However, it is clear from Fig.3
that $\Lambda$ is bounded from above. This upper bound on
$\Lambda$ ($\Lambda_{upper}$) arises due to the
inequality (46), i.e., there will be negative
soft SUSY-breaking squared masses at
the low-energy scale $\mu_{E}=v_{N_{r}}$ if $\Lambda\,>\,\Lambda_{upper}$.
It explicitly confirms the conclusion obtained in \mbox{Section 6}
that the MSMHHC should be an effective intermediate-scale model.
In addition, this upper bound on $\Lambda$
also implies a lower bound on $m_{LH}$, which is consistent with (45).
That is, as $f_{T}$ increases, $\Lambda_{upper}$ decreases according to (46),
and therefore the lower bound on $m_{LH}$ increases according to Fig.3.

    As shown in Fig.3, the prediction of $m_{LH}$ is insensitive to the
choice of $m_{top}$ within \mbox{164 GeV $\leq\;m_{top}\;\leq$ 180 GeV}
unless $\Lambda$ is close to $\Lambda_{upper}$. Therefore, in
\mbox{Sections 8 and 9}, only the case of \mbox{$m_{top}=180$ GeV} will
be studied. Notice that (42) is not applicable to Fig.3 because
(42) is valid only when $m_{LH}$ is comparatively small. The predictions of
Fig.3 do lie outside the linear region described by (42). It will be shown
in Fig.6 of \mbox{Section 8} that the predictions of Fig.3 lie
in the saturated region due to the comparatively large
choice \mbox{$F_{SUSY}=800$ GeV}. Due to saturation, $m_{LH}$ depends on
$f_{T}$ mainly through the renormalization effects, not (42). Therefore,
it explains why $m_{LH}$ predicted by \mbox{$m_{top}=180$ GeV} is slightly
smaller than that predicted by \mbox{$m_{top}=164$ GeV}.

    $\tan\beta_{r}=\frac{v_{2r}}{v_{1r}}$ versus $\Lambda$ is plotted in
Fig.4. Due to the choice \mbox{$\frac{m_{1}}{m_{2}}=2$}, $\tan\beta_{r}$ lies
between \mbox{$\,1.5\,\sim\,3.5\,$} and slightly increases as $\Lambda$
increases. As a rough approximation, $\tan\beta_{r}$ is
obtained from (24) and (25) by turning $f_{T}$ off:
\begin{eqnarray}
&&\tan\beta_{r}=\frac{v_{2r}}{v_{1r}}\,\approx\,
  \sqrt{\frac{m_{1}^{2}+h^{2}v_{N}^{2}}{m_{2}^{2}+h^{2}v_{N}^{2}}}
\end{eqnarray}
Therefore, $\tan\beta_{r}$ is essentially determined by the ratio
$\frac{m_{1}}{m_{2}}$. Different choices of $\frac{m_{1}}{m_{2}}$
will be made in \mbox{Section 9}. In Fig.5, the plot of
$\mu_{r}=h_{r}<\phi_{N_{r}}>$ versus $\Lambda$ is given. $\mu_{r}$
decreases as $\Lambda$ increases and, as expected, $\mu_{r}$ is always
of order \mbox{$F_{SUSY}=800$ GeV}, the soft SUSY-breaking strength.
\section{The Dependence of Low-Energy \mbox{Quantities}
         on {\bf $F_{SUSY}$}}
\hspace{0.8cm}In this section, we would like to study the dependence
on the soft SUSY-breaking strength $F_{SUSY}$ by fixing the soft
SUSY-breaking pattern (i.e., keeping the ratio between any two soft
SUSY-breaking parameters fixed). The dependence on $F_{SUSY}$ is
illustrated with a typical example, where its parameters are chosen as:
\mbox{$\Lambda=500$ TeV}, \mbox{$m_{1}=2\,F_{SUSY}$}, \mbox{$m_{2}=F_{SUSY}$},
\mbox{$m_{12}=\frac{1}{2}\,F_{SUSY}$}, \mbox{$m_{top}$=180 GeV},
and $F_{SUSY}$ is left free. Following the computational procedure
described in \mbox{Section 7}, $m_{LH}$ versus $F_{SUSY}$
is plotted in Fig.6. When $F_{SUSY}$ is small (e.g.,
\mbox{$F_{SUSY}<500$ GeV} for Fig.6), $m_{LH}$ is almost proportional to
$F_{SUSY}$, which has been predicted by (42).  When $F_{SUSY}$ is large
(e.g., \mbox{$F_{SUSY}>500$ GeV} for Fig.6), (42) is no longer valid and
$m_{LH}$ actually saturates. Therefore, $m_{LH}$ depends on $F_{SUSY}$ in a
non-trivial way and, due to the saturation of $m_{LH}$,
the prediction of $m_{LH}$ by the MSMHHC is stable under the variation of
$F_{SUSY}$ when $F_{SUSY}$ is large. On the other hand, $m_{LH}$ is sensitive
to $F_{SUSY}$ when $F_{SUSY}$ is small and the measurement of
$m_{LH}$ can provide us with more knowledge of $F_{SUSY}$ .

    As argued in \mbox{Section 2}, $\mu_{r}$ is induced only through
supersymmetry breaking effects, and therefore the \mbox{$\mu$ problem} is
solved naturally. This argument is realized explicitly by
plotting \mbox{$\mu_{r}=h_{r}<\phi_{N_{r}}>$} versus $F_{SUSY}$ in Fig.7.
Fig.7 shows that $\mu_{r}$ is simply proportional to $F_{SUSY}$ and of order
$F_{SUSY}$, which constitutes the concrete answer to the naturalness problem
of $\mu$.
\section{The Dependence of Low-Energy \mbox{Quantities}
         on \mbox{the Soft SUSY-Breaking Pattern}}
\hspace{0.8cm}In this section, the dependence on the soft SUSY-breaking
pattern is studied. That is, $\Lambda$ and $F_{SUSY}$ are fixed, and
the variation of the soft SUSY-breaking pattern means changing
the ratio between any two soft SUSY-breaking parameters. First,
we are interested in the dependence of the low-energy quantities on
the ratio $\frac{m_{1}}{m_{2}}$ with $\sqrt{m_{1}^{2}+m_{2}^{2}}$ fixed.
According to the parameter space (48) and (49), a typical example is chosen
as: \mbox{$\Lambda=500$ TeV}, \mbox{$\sqrt{m_{1}^{2}+m_{2}^{2}}=1$ TeV},
\mbox{$m_{12}=250$ GeV}, \mbox{$m_{top}$=180 GeV}, and $\frac{m_{1}}{m_{2}}$
is left free. The reason why we choose this pattern is that $\tan\beta_{r}$
is essentially determined by $\frac{m_{1}}{m_{2}}$ and we would like to
know how physical quantities, such as $m_{LH}$, depend on $\tan\beta_{r}$.
Remember that the requirement that low-energy soft SUSY-breaking scalar
squared masses be non-negative leads to $m_{T}\approx m_{2}$, and therefore
$m_{T}=m_{2}$ is assumed in (49). Hence, in the above example, $m_{T}$
actually changes according to $m_{T}=m_{2}$ when $\frac{m_{1}}{m_{2}}$
changes. This understanding of how the soft SUSY-breaking pattern is
varied in the above example is crucial to the interpretation of its results,
Fig.8 and Fig.9. The above example is computed according to the procedure
outlined in \mbox{Section 7}, and $\tan\beta_{r}$ versus
$\frac{m_{1}}{m_{2}}$ is plotted in Fig.8. It's obvious that
\mbox{$\tan\beta_{r}\approx \frac{m_{1}}{m_{2}}$}
is a good approximation to Fig.8, especially when $\frac{m_{1}}{m_{2}}$
is large. However, notice that the computations associated with
\mbox{$\tan\beta_{r}\geq O(10)$} are unsatisfactory from the viewpoint of
phenomenology because all the Yukawa couplings are ignored except for
the top Yukawa coupling in this paper. That is, it no longer makes sense
to ignore the bottom Yukawa coupling when \mbox{$\tan\beta_{r}\geq O(10)$}.
Therefore, only those results with \mbox{$\tan\beta_{r}< O(10)$} in
Fig.8 and Fig.9 are meaningful to phenomenology. Those results with
\mbox{$\tan\beta_{r}\geq O(10)$} needs to be re-computed by including
the bottom Yukawa coupling. Keeping the above discussion in mind,
we still consider the case \mbox{$\tan\beta_{r}\gg 1$} in the following
in order to verify the basic features of Fig.8 and Fig.9. That is, the
verification of their basic features is meant to be a consistency check
only. The approximation \mbox{$\tan\beta_{r}\approx \frac{m_{1}}{m_{2}}$}
is consistent with (51), and we will show it when
\mbox{$\frac{m_{1}}{m_{2}}\gg 1$}. Based on \mbox{$\frac{m_{1}}{m_{2}}\gg 1$}
and \mbox{$\sqrt{m_{1}^{2}+m_{2}^{2}}=$constant},
the vacuum constraint (24) implies
\mbox{$hv_{F_{N}}\propto \frac{1}{\tan\beta_{r}}$}.
With \mbox{$hv_{F_{N}}\propto \frac{1}{\tan\beta_{r}}$}, the vacuum
constraint (25) implies \mbox{$hv_{N}\propto \frac{1}{\tan\beta_{r}}$}.
When \mbox{$\frac{m_{1}}{m_{2}}\gg 1$}, \mbox{$\tan\beta_{r}\gg 1$} according
to Fig.8, and therefore \mbox{$m_{1}\gg hv_{N}$}. With
\mbox{$m_{1}\gg hv_{N}$}, (51) implies that $\tan\beta_{r}$ is of
order $\frac{m_{1}}{m_{2}}$, which is indeed consistent with Fig.8.

    The above argument also leads us to the qualitative dependence of
$m_{LH}$ on $\tan\beta_{r}$. According to the
above argument, \mbox{$hv_{N}\propto \frac{1}{\tan\beta_{r}}$} and
\mbox{$m_{T}=m_{2}\propto \frac{1}{\tan\beta_{r}}$}
when \mbox{$\frac{m_{1}}{m_{2}}\gg 1$}. Referring to (41) and (42),
one therefore has \mbox{$\,\sigma_{1}\propto \frac{1}{\tan^{2}\beta_{r}}\,$}
and \mbox{$\,\sigma_{2}\approx 2\,$} when \mbox{$\frac{m_{1}}{m_{2}}\gg 1$}.
Using (42) as a qualitative approximation to $m_{LH}$, one obtains the
following result which shows the \mbox{$\tan\beta_{r}$-dependence} only:
\begin{eqnarray}
&&m_{LH}\;\propto\;\sqrt{\frac{U_{2}}{\tan^{2}\beta_{r}}\,+\,
        \frac{U_{1}}{\tan^{4}\beta_{r}}}
\end{eqnarray}
where $U_{1}$ and $U_{2}$ are inessential coefficients of order unity.
When \mbox{$\frac{m_{1}}{m_{2}}\gg 1$}, \mbox{$\tan\beta_{r}\gg 1$} and
\mbox{$m_{LH}\propto \frac{1}{\tan\beta_{r}}$} from
(52). To confirm (52),
$m_{LH}$ versus $\tan\beta_{r}$ is plotted in Fig.9 by changing
$\frac{m_{1}}{m_{2}}$. For \mbox{$1<\tan\beta_{r}<11$},
\mbox{ 50 GeV $\leq\,m_{LH}\,\leq$ 250 GeV}. It's clear from Fig.9 that
$m_{LH}$ does decrease as $\tan\beta_{r}$ increases, which is consistent
with (52). The way $m_{LH}$ depends on $\tan\beta_{r}$ has an important
implication: Due to the present experimental lower bound on $m_{LH}$
\cite{Lopez93}, large values of $\tan\beta_{r}$ are not
favored, i.e., $\tan\beta_{r}$ must be bounded from above.
For example, \mbox{$\tan\beta_{r}<9$} in Fig.9 if the
experimental lower bound on $m_{LH}$ is taken to be \mbox{60 GeV}.
However, as mentioned before, this upper bound on $\tan\beta_{r}$ is
inconclusive unless the bottom Yukawa coupling is included.

    Second, the dependence on the soft SUSY-breaking parameter $m_{12}$
is studied. It is illustrated with a typical example whose parameters
are chosen as: \mbox{$\Lambda=500$ TeV}, \mbox{$m_{1}=1.6$ TeV},
\mbox{$m_{2}=800$ GeV}, \mbox{$m_{top}$=180 GeV}, and $m_{12}$ is left free.
The effect of $m_{12}$ is best seen through $m_{LH}$, and therefore
$m_{LH}$ versus $\frac{m_{12}}{m_{2}}$ is plotted in Fig.10. As discussed in
\mbox{Section 2}, the Higgs sector of $\cal L$$_{\Lambda}$ has an
SU(2)$\times$U(1)$\times$U(1) symmetry instead of SU(2)$\times$U(1) in the
limit $m_{12}=0$. Therefore, when $m_{12}=0$, there is one more U(1)
symmetry, and the spontaneous breaking of this additional U(1) symmetry
leads to one more massless Goldstone boson, which is exactly the lightest
Higgs boson. This explains why $m_{LH}=0$ in the limit \mbox{$m_{12}=0$},
and $m_{LH}$ increases linearly with $m_{12}$
\mbox{(when $\frac{m_{12}}{m_{2}}<\frac{1}{3}$)} in Fig.10.
However, when $m_{12}$ is large enough
\mbox{($\frac{m_{12}}{m_{2}}>\frac{1}{3}$ for Fig.10)}, $m_{LH}$
is determined by another mechanism: $m_{12}$ contributes only to
the off-diagonal elements of the Higgs squared-mass matrix, and therefore
$m_{LH}$ decreases with $m_{12}$ when $m_{12}$ is large enough.
In Fig.10, the non-smoothness of the curve around
\mbox{$\frac{m_{12}}{m_{2}}=\frac{1}{3}$} simply reflects that $m_{LH}$ is
determined by different mechanisms for
\mbox{$\frac{m_{12}}{m_{2}}<\frac{1}{3}$}
and \mbox{$\frac{m_{12}}{m_{2}}>\frac{1}{3}$}. Again, due to the experimental
lower bound on $m_{LH}$, $m_{12}$ must be bounded from both below and
above according to Fig.10.
\section{Conclusions}
\hspace{0.8cm}In this paper, the idea of Higgs-Higgs condensation
is proposed, and it is pointed out how three relevant issues can be solved
based on the Higgs-Higgs condensation: the \mbox{$\mu$ problem}, the
possibility of raising the lightest-Higgs mass, and the triviality problem
associated with the Higgs sector. As the first realization of this idea,
the Minimal Supersymmetric Model of Higgs-Higgs Condensation is
constructed and its qualitative features are studied in detail. Finally, in
order to reveal the phenomenological details of the MSMHHC, a systematic
study of its parameter space is made. Obviously, an immediate extension of
this paper is to include the electroweak SU(2)$\times$U(1) gauge
interactions. However, the mere inclusion of gauge interactions may not be
enough to solve the fine-tuning problem associated with the soft
SUSY-breaking
parameters (as discussed in \mbox{Section 6}), and therefore more general
structures of the Higgs non-renormalizable interactions should be considered
in the future. Another relevant issue that must be faced in the future
model-building is that the lack of guiding principles in choosing the
structure of Higgs non-renormalizable interactions will undermine
the predictive power of the model. As pointed out in \mbox{Section 6},
thanks to the requirement of no fine tuning, this fine-tuning problem
actually serves as a guide to the selection of the structure of Higgs
non-renormalizable interactions, and therefore both the fine-tuning problem
and the issue of the arbitrariness in choosing non-renormalizable
interactions may be solved at the same time in future works.

    Finally, let's reveal the physical significance of Higgs-Higgs
condensation from another viewpoint by asking the following question:
"How large can the lightest-Higgs mass be in supersymmetric theories?"
According to the studies of the Non-Minimal SSM \cite{yywu,NMSSM,Moroi92},
the first part of the answer is that the Non-Minimal SSM's are capable of
predicting larger lightest-Higgs mass than the MSSM (or the MTCSSM) is.
The second part of the answer follows from the present paper: Among the
Non-Minimal SSM's, the supersymmetric standard model of Higgs-Higgs
condensation is preferred because it is the only choice free from being
plagued by the triviality problem. Therefore, if the lightest-Higgs mass
predicted by the MSSM is excluded by the future experiments, the
supersymmetric model of Higgs-Higgs condensation seems to be the most
promising candidate.
\section*{Acknowledgement}
\hspace{0.8cm}I would like to thank Professor Mary K. Gaillard
for her support and nice advice.
This work was supported in part by the Director, Office of
Energy Research, Office of High Energy and Nuclear Physics, Division of
High Energy Physics of the U.S. Department of Energy under Contract
DE-AC03-76SF00098 and in part by the National Science Foundation under
grant PHY90-21139.
\pagebreak

\pagebreak
\clearpage
\hspace{1.6in} FIGURE CAPTIONS

\vskip 0.5in
Fig.1: The supergraph contributing to the induced kinetic term \linebreak
\mbox{$\Sigma_{N}\int d^{4}\theta\,N^{\dagger}N
[1+(m_{1}^{2}+m_{2}^{2})\theta^{2}\bar{\theta}^{2}]$}
\mbox{for the Higgs chiral superfield $N$.}

\vskip 0.5in
Fig.2: A plot of the allowed range of the soft SUSY-breaking parameter
\mbox{$B$ (GeV)}:
\mbox{$B_{lower}\leq B <B_{upper}$} versus the strength of soft SUSY breaking
\mbox{$F_{SUSY}$ (GeV)}, where the dotted (solid) line corresponds to
$B_{upper}$ ($B_{lower}$).
According to (35), the parameters are chosen as: \mbox{$\Lambda=2.5\times
10^{4}$ TeV},
\mbox{$f_{T}=1.6\,$}, \mbox{$m_{1}=2\,F_{SUSY}$},
\mbox{$m_{2}=m_{T}=F_{SUSY}$},
\mbox{$m_{M}=\sqrt{5}\,F_{SUSY}$}, \mbox{$m_{12}=\frac{1}{2}\,F_{SUSY}$}. $B$
and $F_{SUSY}$
are left free.

\vskip 0.5in
Fig.3: Plots of the lightest-Higgs mass \mbox{$m_{LH}$ (GeV)} versus the
cut-off \mbox{$\Lambda$ (GeV)}
for two different choices of the top-quark mass: \mbox{$m_{top}=164$ GeV} and
\mbox{$m_{top}=180$ GeV}.
According to (48) and (49), the parameters are chosen as: \mbox{$F_{SUSY}=800$
GeV},
\mbox{$m_{1}=2\,F_{SUSY}$}, \mbox{$m_{2}=F_{SUSY}$},
\mbox{$m_{12}=\frac{1}{2}\,F_{SUSY}$},
\mbox{$m_{top}$=164 or 180 GeV}, and $\Lambda$ is left free.

\vskip 0.5in
Fig.4: Plots of $\tan\beta_{r}=\frac{v_{2r}}{v_{1r}}$ versus the cut-off
\mbox{$\Lambda$ (GeV)}
for two different choices of the top-quark mass: \mbox{$m_{top}=164$ GeV} and
\mbox{$m_{top}=180$ GeV}.
According to (48) and (49), the parameters are chosen as: \mbox{$F_{SUSY}=800$
GeV},
\mbox{$m_{1}=2\,F_{SUSY}$}, \mbox{$m_{2}=F_{SUSY}$},
\mbox{$m_{12}=\frac{1}{2}\,F_{SUSY}$},
\mbox{$m_{top}$=164 or 180 GeV}, and $\Lambda$ is left free.

\vskip 0.3in
Fig.5: Plots of the \mbox{effective $\mu_{r}=h_{r}<\phi_{N_{r}}>$ (GeV)} versus
the cut-off \mbox{$\Lambda$ (GeV)}
for two different choices of the top-quark mass: \mbox{$m_{top}=164$ GeV} and
\mbox{$m_{top}=180$ GeV}.
According to (48) and (49), the parameters are chosen as: \mbox{$F_{SUSY}=800$
GeV},
\mbox{$m_{1}=2\,F_{SUSY}$}, \mbox{$m_{2}=F_{SUSY}$},
\mbox{$m_{12}=\frac{1}{2}\,F_{SUSY}$},
\mbox{$m_{top}$=164 or 180 GeV}, and $\Lambda$ is left free.

\vskip 0.3in
Fig.6: A plot of the lightest-Higgs mass \mbox{$m_{LH}$ (GeV)} versus the
strength of soft SUSY breaking
\mbox{$F_{SUSY}$ (GeV)}. According to (48) and (49), the parameters are chosen
as: \mbox{$\Lambda=500$ TeV},
\mbox{$m_{1}=2\,F_{SUSY}$}, \mbox{$m_{2}=F_{SUSY}$},
\mbox{$m_{12}=\frac{1}{2}\,F_{SUSY}$},
\mbox{the top-quark mass $m_{top}$=180 GeV}, and $F_{SUSY}$ is left free.

\vskip 0.3in
Fig.7: A plot of the \mbox{effective $\mu_{r}=h_{r}<\phi_{N_{r}}>$ (GeV)}
versus the strength of soft SUSY breaking
\mbox{$F_{SUSY}$ (GeV)}. According to (48) and (49), the parameters are chosen
as: \mbox{$\Lambda=500$ TeV},
\mbox{$m_{1}=2\,F_{SUSY}$}, \mbox{$m_{2}=F_{SUSY}$},
\mbox{$m_{12}=\frac{1}{2}\,F_{SUSY}$},
\mbox{the top-quark mass $m_{top}$=180 GeV}, and $F_{SUSY}$ is left free.

\vskip 0.5in
Fig.8: A plot of $\tan\beta_{r}=\frac{v_{2r}}{v_{1r}}$ versus
$\frac{m_{1}}{m_{2}}$.
According to (48) and (49), the parameters are chosen as: \mbox{$\Lambda=500$
TeV},
\mbox{$\sqrt{m_{1}^{2}+m_{2}^{2}}=1$ TeV}, \mbox{$m_{12}=250$ GeV},
\mbox{the top-quark mass $m_{top}$=180 GeV}, and $\frac{m_{1}}{m_{2}}$ is left
free.

\vskip 0.5in
Fig.9: A plot of the lightest-Higgs mass \mbox{$m_{LH}$ (GeV)} versus
$\tan\beta_{r}=\frac{v_{2r}}{v_{1r}}$.
According to (48) and (49), the parameters are chosen as: \mbox{$\Lambda=500$
TeV},
\mbox{$\sqrt{m_{1}^{2}+m_{2}^{2}}=1$ TeV}, \mbox{$m_{12}=250$ GeV},
\mbox{the top-quark mass $m_{top}$=180 GeV}, and $\frac{m_{1}}{m_{2}}$ is left
free.

\vskip 0.5in
Fig.10: A plot of the lightest-Higgs mass \mbox{$m_{LH}$ (GeV)} versus
$\frac{m_{12}}{m_{2}}$.
According to (48) and (49), the parameters are chosen as: \mbox{$\Lambda=500$
TeV}, \mbox{$m_{1}=1.6$ TeV},
\mbox{$m_{2}=800$ GeV}, \mbox{the top-quark mass $m_{top}$=180 GeV}, and
$m_{12}$ is left free.


\begin{thebibliography}{99}

\bibitem[1]{mu-termN} \mbox{H.P. Nilles,
Phys. Rep. {\bf 110}, 1 (1984).}
\linebreak  \mbox{J.E. Kim and H.P. Nilles,
\mbox{Phys. Lett.} {\bf 138B}, 453 (1984).}

\bibitem[2]{mu-termH} J.A. Harvey, hep-th/9411097 (1994).

\bibitem[3]{Haber91}  H.E. Haber and R. Hempfling,
\mbox{Phys. Rev. Lett.} {\bf 66}, 1815 (1991).

\bibitem[4]{Haber92}  H.E. Haber and R. Hempfling, SCIPP 91/33 (1992).

\bibitem[5]{HiggsHunter} {\em The Higgs Hunter's Guide} by
\mbox{J.F. Gunion et al.} (Addison-Wesley, MA, 1990), pp.267-322.

\bibitem[6]{Callaway} D.J.E. Callaway,
Phys. Rep. {\bf 167}, 241 (1988).

\bibitem[7]{Binetruy92} P. Bin\'{e}truy and C.A. Savoy,
\mbox{Phys. Lett.} {\bf 277B}, 453 (1992).

\bibitem[8]{yywu}  Y.-Y. Wu, \mbox{Phys. Rev. {\bf D 51}}, 5276 (1995).

\bibitem[9]{NMSSM} \mbox{J. Ellis et al.,
\mbox{Phys. Rev. {\bf D 39}}, 844 (1989).}
\linebreak          \mbox{M. Drees,
\mbox{Int. J. Mod. Phys. {\bf A4}}, 288 (1993).}
\linebreak          \mbox{L.E. Ibanez and J. Mas,
\mbox{Nucl. Phys. {\bf B286}}, 107 (1987).}
\linebreak    \mbox{J.P. Derendinger and C. Savoy,
\mbox{Nucl. Phys. {\bf B237}}, 307 (1984).}
\linebreak  \mbox{J.M. Frere, D.R.T. Jones and S. Raby,
\mbox{Nucl. Phys. {\bf B222}}, 11 (1983).}
\linebreak  \mbox{H.P. Nilles, M. Srednicki and D. Wyler,
\mbox{Phys. Lett. {\bf 120B}}, 346 (1983).}

\bibitem[10]{Clark90} T.E. Clark, S.T. Love and W.A. Bardeen,
\mbox{Phys. Lett.} {\bf 237B}, 235 (1990).

\bibitem[11]{Carena92} M. Carena et al.,
\mbox{Nucl. Phys. {\bf B369}}, 33 (1992).

\bibitem[12]{Binetruy94} P. Bin\'{e}truy, E.A. Dudas and F. Pillon,
\mbox{Nucl. Phys. {\bf B415}}, 175 (1994).

\bibitem[13]{Buchmuller84} W. Buchm\"{u}ller and U. Ellwanger,
\mbox{Nucl. Phys. {\bf B245}}, 237 (1984).

\bibitem[14]{Polchinski82} J. Polchinski and L. Susskind,
\mbox{Phys. Rev. {\bf D 26}}, 3661 (1982).

\bibitem[15]{PQ} R.D. Peccei and H.R. Quinn,
\mbox{Phys. Rev. Lett. {\bf 38}}, 1440 (1977).

\bibitem[16]{Kim91} J.E. Kim and H.P. Nilles,
\mbox{Phys. Lett. {\bf 263B}}, 79 (1991).

\bibitem[17]{Chun92} E.J. Chun, J.E. Kim and H.P. Nilles,
\mbox{Nucl. Phys. {\bf B370}}, 105 (1992).

\bibitem[18]{Chun94} E.J. Chun, hep-ph/9411290 (1994).

\bibitem[19]{Guidice88} \mbox{G.F. Guidice and A. Masiero,
\mbox{Phys. Lett. {\bf 206B}}, 480 (1988).}
\linebreak    \mbox{J.A. Casas and C. Munoz,
\mbox{Phys. Lett. {\bf 306B}}, 288 (1993).}
\linebreak    \mbox{A. Brignole, L.E. Ibanez and C. Munoz,
\mbox{Nucl. Phys. {\bf B422}}, 125 (1994).}
\linebreak    \mbox{I. Antoniadis et al.,
hep-th/9405254 (1994).}

\bibitem[20]{Moroi92} \mbox{T. Moroi and Y. Okada,
\mbox{Phys. Lett. {\bf 295B}}, 73 (1992).}
\linebreak    \mbox{T. Elliott, S.F. King and P.L. White,
\mbox{Phys. Rev. {\bf D 49}}, 2435 (1994).}
\linebreak    \mbox{W.T.A. ter Veldhuis,
PURD-TH-92-11 (1992).}

\bibitem[21]{Lindner93} M. Lindner,
\mbox{Int. J. Mod. Phys. {\bf A8}}, 2167 (1993).

\bibitem[22]{Wess} {\em Supersymmetry and Supergravity} by
\mbox{J. Wess and J. Bagger}
(Princeton University Press, NJ, 1992).

\bibitem[23]{Buchmuller82} W. Buchm\"{u}ller and S.T. Love,
\mbox{Nucl. Phys. {\bf B204}}, 213 (1982).

\bibitem[24]{Coleman73} \mbox{S. Coleman and E. Weinberg,
\mbox{Phys. Rev. {\bf D 7}}, 1883 (1973).}
\linebreak    \mbox{R. Barbieri and S. Cecotti,
\mbox{Z. Phys. {\bf C17}}, 183 (1983).}
\linebreak    \mbox{P. Bin\'{e}truy et al.,
\mbox{Phys. Rev. {\bf D 37}}, 2633 (1988).}

\bibitem[25]{Falck86} N.K. Falck,
\mbox{Z. Phys. {\bf C30}}, 247 (1986).

\bibitem[26]{Suzuki90} \mbox{W. Bardeen,
\mbox{Fermilab preprint FERMILAB-CONF-90/269-T} (1990).}
\linebreak    \mbox{M. Suzuki,
\mbox{Mod. Phys. Lett. {\bf A5}}, 1205 (1990).}
\linebreak    \mbox{A. Hasenfratz et al.,
\mbox{Nucl. Phys. {\bf B365}}, 79 (1991).}

\bibitem[27]{Lopez93} \mbox{J.L. Lopez et al.,
\mbox{Phys. Lett. {\bf 306B}}, 73 (1993).}
\linebreak  \mbox{{\em Research Directions For The Decade} edited by
\mbox{E.L. Berger}} \mbox{(World Scientific, Singapore, 1990).}

\end{thebibliography}
\end{document}